\documentclass[11pt]{article}
\pdfoutput=1
\usepackage{jcappub,natbib,float,comment,bm,placeins} 

\def\be{\begin{equation}}
\def\ee{\end{equation}}
\def\ba#1\ea{\begin{align}#1\end{align}}

\renewcommand{\v}[1]{\mathbf{#1}}
\newcommand{\vx}{\v{x}}

\newcommand{\refeq}[1]{Eq.~(\ref{eq:#1})}          
          
\newcommand{\reffig}[1]{figure~\ref{fig:#1}} 
\newcommand{\reffigs}[2]{figures~(\ref{fig:#1})--(\ref{fig:#2})}
\newcommand{\refFig}[1]{Figure~\ref{fig:#1}}
\newcommand{\refFigs}[2]{Figures~(\ref{fig:#1})--(\ref{fig:#2})}          
\newcommand{\refsec}[1]{section~\ref{sec:#1}}          
\newcommand{\refSec}[1]{Section~\ref{sec:#1}}          
\newcommand{\refapp}[1]{Appendix~\ref{app:#1}}
\newcommand{\reftab}[1]{table~\ref{table:#1}}

\newcommand{\Om}{\Omega_m}
\newcommand{\Ob}{\Omega_b}

\newcommand{\OL}{\Omega_\Lambda}

\renewcommand{\d}{\delta}

\newcommand{\dr}{\delta_\rho}

\def\Mpch{\,h^{-1}{\rm Mpc}}
\newcommand{\bie}[1]{b_#1^E}

\newcommand{\bir}[1]{b_#1^{\rm rel}}
\newcommand{\ESP}{\mathrm{ESP}}

\newcommand{\<}{\left\langle}
\renewcommand{\>}{\right\rangle}

\title{Large-scale assembly bias of dark matter halos}

\author[a]{Titouan Lazeyras,}
\author[a]{Marcello Musso,}
\author[a]{Fabian~Schmidt}
\affiliation[a]{Max-Planck-Institut f\"ur Astrophysik, Karl-Schwarzschild-Str. 1, 85748 Garching, Germany}
\emailAdd{titouan@mpa-garching.mpg.de, mmusso@sas.upenn.edu, fabians@mpa-garching.mpg.de}
\abstract{We present precise measurements of the assembly bias of dark matter halos, i.e. the dependence of halo bias on other properties than the mass, using curved ``separate universe'' N-body simulations which effectively incorporate an infinite-wavelength matter overdensity into the background density.  This method measures the LIMD (local-in-matter-density) bias parameters $b_n$ in the large-scale limit.  
We focus on the dependence of the first two Eulerian biases $b^E_1$ and $b^E_2$ on four halo properties: the concentration, spin, mass accretion rate, and ellipticity.  We quantitatively compare our results with previous works in which assembly bias was measured on fairly small scales. Despite this difference, our findings are in good agreement with previous results. We also look at the joint dependence of bias on two halo properties in addition to the mass. Finally, using the excursion set peaks model, we attempt to shed new insights on how assembly bias arises in this analytical model.}

\keywords{dark matter halos, bias, galaxy clustering, cluster counts}
\begin{document}
\maketitle
\flushbottom

%%%%%%%%%%%%%%%%%%%%%%%%%%%%%%%%%%%%%%%%%%%%%%%%%%%%%%%%%%%%%%%%%%%%%%%%%%%%%
%%%%%%%%%%%%%%%%%%%%%%%%%%%%%%%%%%%%%%%%%%%%%%%%%%%%%%%%%%%%%%%%%%%%%%%%%%%%%
\section{Introduction}
\label{sec:intro}

The large-scale distribution of dark matter halos is one of the key
ingredients of the theoretical description of large-scale structure (LSS).  
Since most observed tracers of LSS, such as galaxies, reside in halos,
the statistics of halos determine those of galaxies on large scales.  
In the context of perturbation theory, the statistics of halos are written in terms of bias parameters 
multiplying operators constructed out of the matter density field (see \cite{Desjacques:2016} for a recent review).  
The most well-studied and phenomenologically most important
bias parameters on large scales are those multiplying powers of
the matter density field, i.e. 
\be
\d_h(\vx,\tau) \supset b^E_1(\tau) \dr(\vx,\tau) + \frac12 b^E_2(\tau) \dr^2(\vx,\tau)
+ \frac16 b^E_3(\tau) \dr^3(\vx,\tau) + \cdots\,,
\label{eq:localbias}
\ee
where $\d_h$ is the fractional number density perturbation of a given
halo sample,
while $\dr$ is the matter density perturbation. More precisely,
the powers of $\dr$ should be understood as renormalized operators
\cite{mcdonald,assassi/etal,PBSpaper}.  
Following the terminology proposed in \cite{Desjacques:2016}, we refer to the $b^E_n$ as local-in-matter-density, or LIMD, bias parameters. Here and in the following the superscript $E$ stands for Eulerian and means that we focus on the bias parameters at late time.  

These bias parameters were commonly thought to depend only on the redshift and mass of the considered halo population, implying that the clustering of dark matter halos is unaffected by the halo environment. This is a central result from basic analytical models for the clustering of matter, such as the excursion set with uncorrelated steps and a constant barrier (\cite{Bond:1990, Lacey:1993, mo/white:1996}). Furthermore, it is a central assumption in several (semi-)analytical models for halo and galaxy statistics, such as  the halo occupation distribution (HOD) model (e.g. \cite{Kauffmann:1995, Berlind:2002, Yang:2002, Zheng:2004} and references therein). In the last decade however, several studies showed that such a model of halo biasing is too simplistic and that halo bias, and more generally halo formation, depend on several other halo properties affected by the halo environment (see for instance \cite{Sheth:2004, gao:2005, Gao:2006, Wechsler:2005, Jing:2006, Croton:2006, Angulo:2007, Dalal:2008, Faltenbacher:2009, Sunayama:2015, Miyatake:2015, More:2016} and references therein). This phenomenon is now known as \textit{assembly bias}. 

Using the so-called marked correlations technique, Ref.~\cite{Sheth:2004} were the first to show that halo formation depends on the halo environment, which provided the first indirect evidence of assembly bias. Shortly after, Ref.~\cite{gao:2005} presented the first direct measurements of this effect. They looked at the dependence of halo bias on the halo formation time, parametrized as the redshift at which the main halo progenitor had assembled half of the total final halo mass. Using the upper and lower tails of the formation time distribution to create two subsamples of their total population, they found clear evidence for assembly bias, with older halos being more clustered than average and younger halos less clustered. Later on, Ref.~\cite{Wechsler:2005} (referred to as W06 in the following) studied assembly bias as a function of concentration. Since then, numerous works, using numerical simulations, have studied and found assembly bias as a function of various halo properties in addition to formation time and concentration, such as spin, shape, substructure content or mass accretion rate. One thing that quickly became clear is that possible correlations between two halo properties are not sufficient to explain the change in clustering observed with respect to these quantities. The most stringent example for that is the fact that there exists a mass range (roughly between $M_\star$ and $5M_\star$, where $M_\star$ is the typical mass of halos that just collapsed today) where older halos are more clustered but more concentrated halos are less clustered (\cite{Gao:2006, Jing:2006}) although the halo formation time--concentration relation indicates that older halos are more concentrated (\cite{Navarro:1996, Wechsler:2001, Neto:2007}). In this context, \cite{Croton:2006} suggested that there may be a more fundamental parameter governing halo clustering. Since then, many papers studied the dependence of halo properties as a function of their environment and formation time (see \cite{Borzyszkowski:2016, Lee:2016, Ludlow:2016} for recent works). Despite the efforts in this direction, a fully consistent picture explaining the physical mechanisms behind assembly bias and which halo properties govern it is still lacking to this day.

Halo assembly bias is not only studied in simulations. Indeed, several studies claimed to have observed assembly bias on galaxy scales (see for instance \cite{Yang:2005, Tinker:2012, Wang:2013}). However, some of these claims were re-investigated by further studies without clear evidence for assembly bias (\cite{Lin:2015}). More recently, \cite{Miyatake:2015, More:2016} presented the first evidence for assembly bias on galaxy cluster scales, using the mean projected cluster-centric distance of member galaxies as a proxy for galactic concentration. The level of assembly bias found by these studies however significantly exceeds the level expected in $\Lambda$CDM and was hence in turn re-investigated by \cite{Zu:2016} who concluded that the signal found in \cite{Miyatake:2015, More:2016} was due to projection effects. Their results for assembly bias on cluster scales are consistent with zero. Finally, \cite{Zentner:2016} used the so-called decorated HOD to put constraints on the level of assembly bias in SDSS DR7 galaxy clustering. While they did not give quantitative results for assembly bias, their study suggests that it is indeed present, especially near the lower luminosity threshold of the sample they study.

Assembly bias is a crucial ingredient in models of the galaxy-halo connection, such as HOD and abundance matching techniques, since a given galaxy sample could preferentially reside in halos with particular properties. Ref. \cite{Zentner:2013} estimated the potential for assembly bias to induce systematic errors in inferred halo occupation statistics. Since then, several studies have introduced improvements of the HOD model (the so-called decorated HOD, \cite{Hearin:2015}) and abundance matching techniques \cite{Chaves-Montero:2015, Lehmann:2015} to include assembly bias. It is important to emphasize however, that the complete \emph{perturbative} bias expansion automatically takes into account assembly bias if all necessary terms are included (see section~9.2 of \cite{Desjacques:2016}).

The goal of this paper is to measure halo assembly bias in $b^E_1$ and $b^E_2$  from simulations using a recently introduced technique, the \textit{separate universe simulations} \citep{lemaitre:1933,sirko:2005,baldauf/etal:2011,li/hu/takada:2014,Wagner:2014}.  This strictly yields the bias parameters in the large-scale limit, i.e. without nonlinear corrections.  In the separate universe approach, a long-wavelength density
perturbation is included in an N-body simulation by changing the 
cosmological parameters, in particular $\Om,\,\OL,\,\Omega_K$ and $H_0$, 
from their fiducial values, and running the simulation to a different
scale factor.  As argued in \cite{baldauf/etal:11,jeong/etal,PBSpaper}, 
the (renormalized) LIMD bias parameters defined in \refeq{localbias}
correspond to the response of the halo abundance, $\bar n_h$, to a long-wavelength
density perturbation, equivalent to a change in the background density, $\bar\rho$,
\be
b^E_{n}(M) = \frac{\bar\rho^{\hskip 1pt n}}{\bar{n}_h(M)} \frac{\partial^n \bar{n}_h(M)}{\partial\bar\rho^{\hskip 1pt n}}\, .
\label{eq:bnsepuniv}
\ee
This can be understood as an exact formulation of the peak-background split (PBS) \cite{kaiser:1984,mo/white:1996}.  Thus, the $b^E_n$ can be measured through the mass function of halos in a suite of separate universe simulations.

The change in the background density is expected to also affect the distribution of other halo properties, such as the halo concentration, formation time, and so on. Hence, by further binning halos in a given mass range $[M, M+{\rm d}M]$ with respect to another quantity, one can measure assembly bias in the same fashion as the standard LIMD bias parameters, with \refeq{bnsepuniv} becoming
\be
b^E_{n}(M,p) = \frac{\bar\rho^{\hskip 1pt n}}{\bar{n}_h(M,p)} \frac{\partial^n \bar{n}_h(M,p)}{\partial\bar\rho^{\hskip 1pt n}}\, ,
\label{eq:asblybnsepuniv}
\ee
where $p$ denotes any halo property other than its mass. In this work, we focus on four halo properties: the concentration, the spin, the shape and the average mass accretion rate on a redshift interval $\Delta z =0.5$.  The same technique was applied very recently by \cite{Paranjape:2016} to measure large-scale halo assembly bias with respect to concentration.
We then attempt to reproduce the assembly bias in a property $p$ by using the assembly bias measured for another property $\tilde p$, and using the mean relation $p(\tilde p)$ between these two quantities. 
We also measure assembly bias with respect to two halo properties, $b^E_n(M,p_1,p_2)$.  

Assembly bias naturally arises in analytical models that go beyond the Markovian excursion set with a constant barrier (e.g. \cite{Zentner:2006, Musso:2012}). Hence we also use the so-called excursion set peaks approach (hereafter ESP) \cite{Paranjape:2012, Paranjape:2013} to investigate how assembly bias emerges in this analytical model of halo clustering, and to compare with the measurements. 

We adopt the same fiducial cosmology as in \cite{Lazeyras:2015}, i.e. a flat $\Lambda {\rm CDM}$ cosmology with $\Om = 0.27$, $h = 0.7$, $\Ob h^2=0.023$ and $\mathcal{A}_s = 2.2 \cdot 10^{-9}$. This paper is organised as follows: in \refsec{theory} we review some aspect of the ESP and show how assembly bias appears in this model. In \refsec{bsep} we describe our simulations and how to measure assembly bias from them. We present and discuss our results in \refsec{results}. We conclude in \refsec{concl}.

%%%%%%%%%%%%%%%%%%%%%%%%%%%%%%%%%%%%%%%%%%%%%%%%%%%%%%%%%%%%%%%%%%%%%%%%%%%%%
%%%%%%%%%%%%%%%%%%%%%%%%%%%%%%%%%%%%%%%%%%%%%%%%%%%%%%%%%%%%%%%%%%%%%%%%%%%%%
\section{Theory predictions}
\label{sec:theory}

The objective of this section is to give a qualitative theoretical understanding of some aspects of assembly bias. To do so, we use the excursion set peaks (ESP) model described in \cite{Paranjape:2012}, \cite{Paranjape:2013}. The model assumes that halos of a given mass $M$ at redshift $z$ are in one-to-one correspondence with peaks of critical height in the linear density field smoothed with a spherical top-hat filter of volume $V=4\pi R_\mathrm{TH}^3/3=M/\bar\rho$. The height of the peak needs to match the critical overdensity
\begin{equation}
  B_\ESP= \delta_c/D(z)+\beta\,\sigma(M)\,,
\label{eq:barrier}
\end{equation}
where $\delta_c=1.686$ is the critical overdensity for spherical collapse, $\sigma(M)$ the square root of the variance of the overdensity field $\delta$ smoothed on some scale corresponding to mass $M$, and $D(z)$ is the linear growth of matter perturbations\footnote{Importantly, upcrossing of the ESP barrier should be thought of as upcrossing of the spherical collapse barrier $\delta_c$ by the process $\delta(\sigma)-\beta\sigma$, which is linear in the density field (see for instance \cite{Musso:2014}). Therefore, it is only $\delta_c$ that should be divided by $D(z)$ to account for linear evolution. However, the term $\beta\sigma$ in $B_\ESP$ still has some redshift dependence because $\sigma=\sigma(M(z))$, which will become important later. We thank Ravi Sheth for pointing this out.}, normalized so that $D(0)=1$. In addition, the model requires that the height of the peak point becomes subcritical when smoothed on slightly larger scales.

The coefficient $\beta$ is an empirical effective parameter that describes the scatter of protohalo densities around $\delta_c$ measured in simulations \cite{Robertson:2008}. As such, it parametrizes our ignorance of how the model of collapse depends on additional variables other than the peak overdensity. Since the first corrections to spherical collapse come from the tidal shear, $\beta$ can be thought of as quantifying the amount of shear acting on the initial protohalo patch (\cite{Castorina:2016}). In the ellipsoidal collapse approximation \cite{Bond:1993}, the tidal field deforms initially spherical perturbations into ellipsoids, and the halo forms when the longest axis (having the slowest infall) recollapses. Since one has to wait for the infall of the slowest axis, the effect of tidal shear slows down collapse.\footnote{See however \cite{Monaco:1997,Pace:2014} for a different interpretation of the role of shear in halo formation} Equivalently, sheared initial overdensities need to be larger than unsheared ones in order to form halos at the same time. The stochastic character of the shear field leads to a stochastic barrier with a linear dependence in $\sigma(M)$ like \refeq{barrier} \cite{Sheth:1999}. The ESP model assumes that $\beta$ is a stochastic variable with mean $\langle\beta\rangle=0.5$ and variance $\langle\beta^2\rangle-\langle\beta\rangle^2=0.25$, which follows a lognormal distribution. These values are simply fitted to reproduce the mass function from N-body simulations, and there is no strong direct evidence in favour of a lognormal distribution (other than that $\beta$ should be positive). However, the predicted linear and quadratic bias coefficients are also in good agreement with simulations results, which provides a non-trivial check of the consistency of the model.

\begin{figure}
\centering
\includegraphics[scale=0.5]{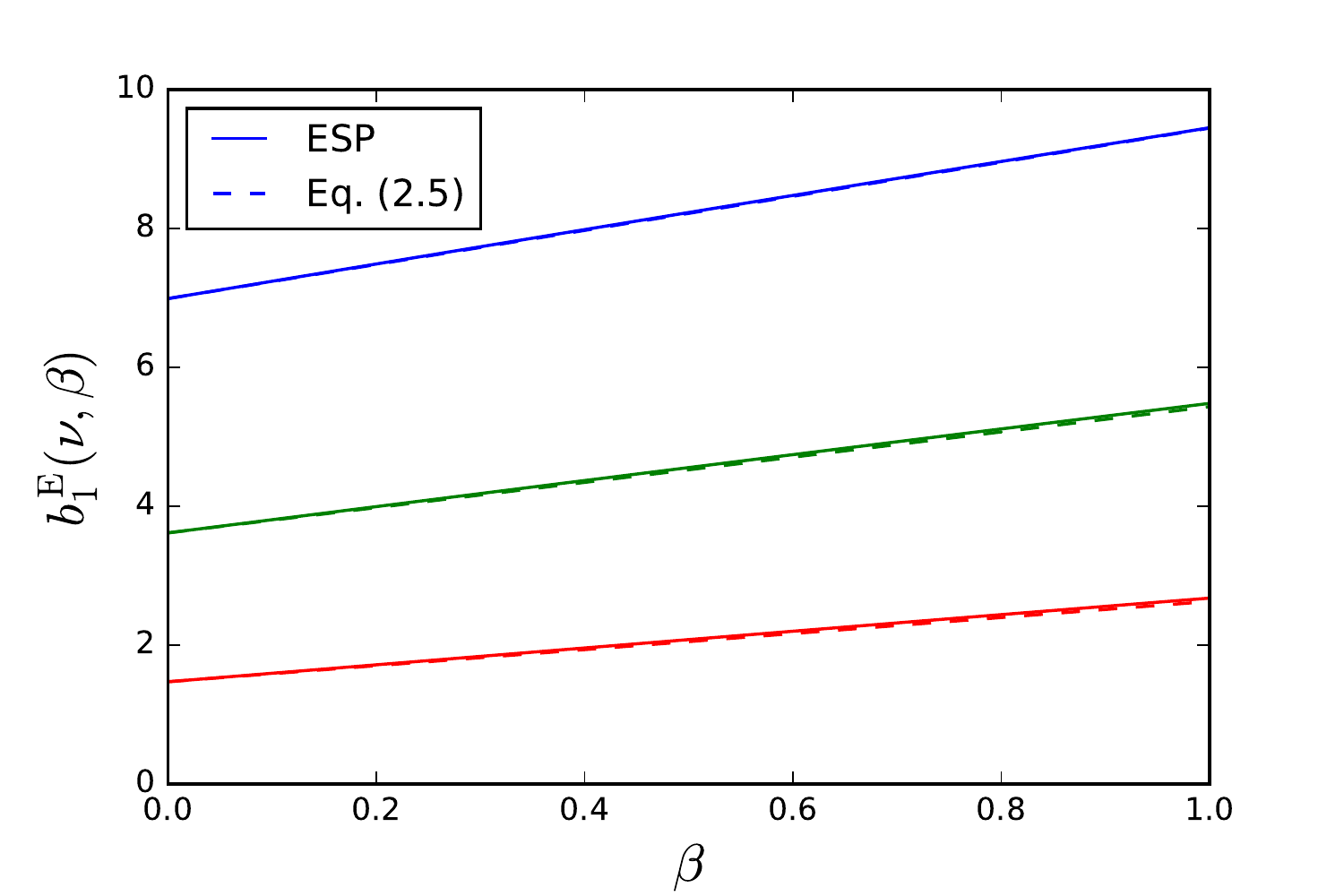}
\includegraphics[scale=0.5]{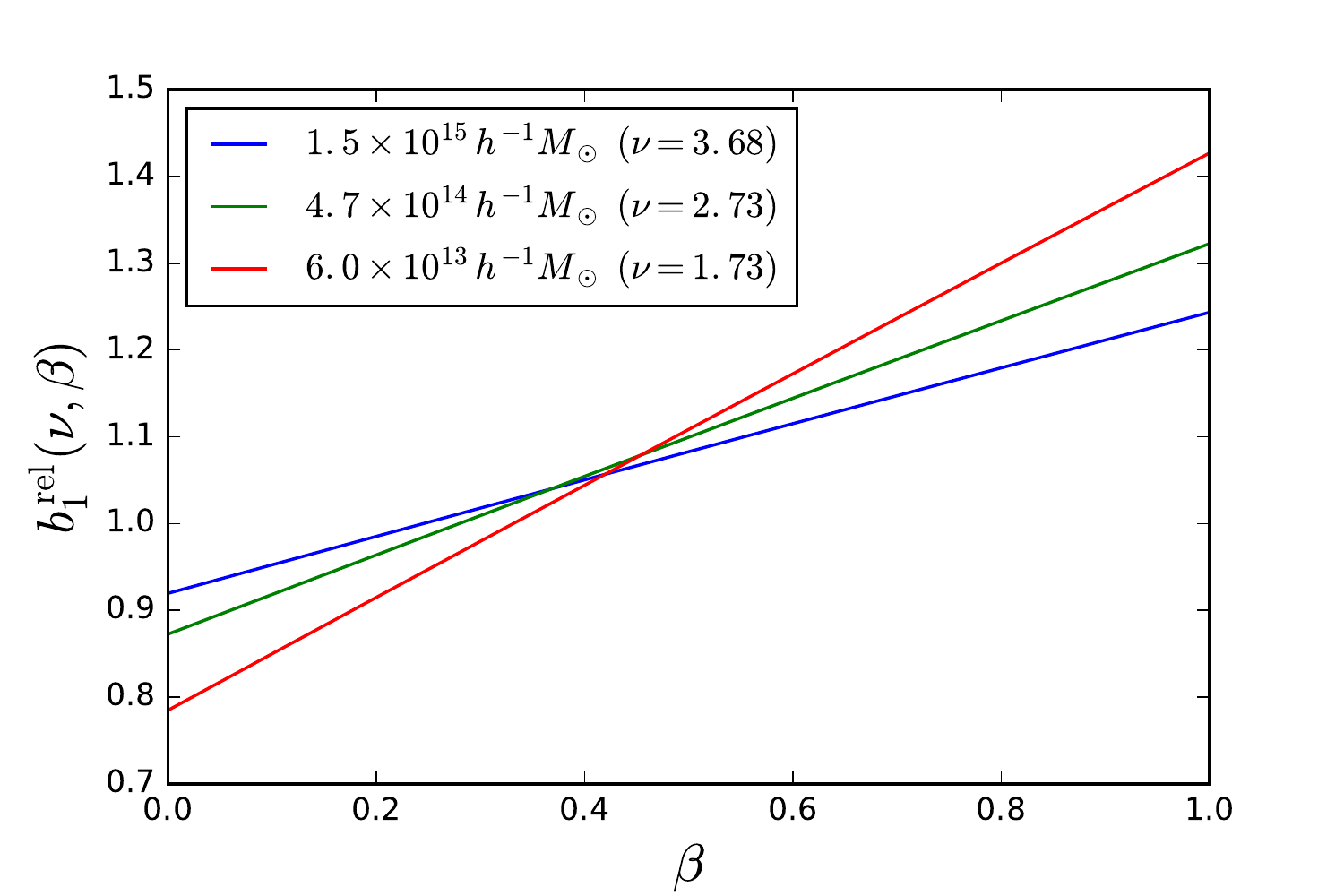}
\caption{$\bie{1}$ and $\bir{1}\equiv b^E_1(\nu,\beta)/b^E_1(\nu)$ as a function of the  $\beta$ parameter, for three different halo masses. Although $\beta$ simply quantifies the scatter of the initial density protohalos, and has no direct physical meaning, it is usually thought to correlate with the amount of tidal shear on the protohalo patch. In the left panel, dashed lines show the linear approximation $\bie{1}(\nu,\beta)=\bie{1}(\nu,0)+(2/3)\beta\nu$.}
\label{fig:b1beta}
\end{figure}

The mean abundance of halos as a function of $\nu=\delta_c/\sigma(M)$ is proportional to the multiplicity function
\begin{equation}
  f(\nu,\beta) = \frac{V}{V_*}
  \frac{e^{-(\nu+\beta)^2/2}}{\sqrt{2\pi}}
  \int_{\gamma\beta}^\infty \!\!\! {\rm d} x 
  \frac{x-\gamma\beta}{\gamma\nu}F(x)
  \mathcal{N}\left(x-\gamma(\nu+\beta), \sqrt{1-\gamma^2}\right),
\label{eq:fESP}
\end{equation}
where $x \equiv -\nabla^2\delta/\langle(\nabla^2\delta)^2\rangle -\gamma\beta$ is the normalized peak curvature, $\gamma\equiv \langle x\delta\rangle/\sigma$ is its cross-correlation with the normalized height, $F(x)$ is the peak curvature function (Eq. (A15) of \cite{Bardeen:1985}) and $\mathcal{N}(X,\sigma)$ denotes a Gaussian distribution for $X$ with zero mean and variance $\sigma^2$.  The peak volume $V_*$ is defined as $V_*\equiv (6\pi\sigma_{1\mathrm{G}}^2/\sigma_{2\mathrm{G}}^2)^{3/2}$, where $\sigma_{1\mathrm{G}}^2$ and $\sigma_{1\mathrm{G}}^2$ are the variances of the gradient and of the Hessian of the density field respectively, smoothed with a Gaussian filter on a scale $R_\mathrm{G}\simeq 0.46 R_\mathrm{TH}$. The large-scale Lagrangian bias parameters at fixed $\beta$ can then be obtained by differentiating $(d\nu/dM) f(\nu,\beta)\propto \nu f(\nu,\beta)$:
\begin{equation}
  b_n^L(\nu,\beta) = 
  \frac{(-1)^n}{\nu f(\nu,\beta)\sigma^n}
  \frac{\partial^n\left[\nu f(\nu,\beta)\right]}{\partial\nu^n}\,,
\label{eq:biastheor}
\end{equation}
which is equivalent to crosscorrelating with Hermite polynomials as defined in \cite{Paranjape:2013}, while the marginalized bias parameters are obtained by differentiating the marginalized multiplicity $\nu f(\nu) = \int {\rm d}\beta p(\beta)\nu f(\nu,\beta)$ as
\begin{equation}
  b_n^L(\nu) = \frac{(-1)^n}{\nu f(\nu)\sigma^{n}}
  \frac{\partial^n\left[\nu f(\nu)\right]}{\partial\nu^n}\,.
\end{equation}
The dependence of the absolute and relative Eulerian bias parameter $b^E_1(\nu\beta)$ and $\bir{1} \equiv b^E_1(\nu,\beta)/b^E_1(\nu)$ on $\beta$ is shown in \reffig{b1beta}. For all values of $\nu$, that is for all masses, bias grows nearly linearly with $\beta$, that is halos with large $\beta$ are more biased. For practical purposes, the dependence is well approximated by
\begin{equation}
   b_1^E(\nu,\beta) =
   1 + b_1^L(\nu,\beta) 
   \simeq  b_1^E(\nu,0) + \frac{2}{3}\beta\nu\,,
\end{equation}
as the dashed lines in \reffig{b1beta} demonstrate.

\begin{figure}
\centering
\includegraphics[scale=0.7]{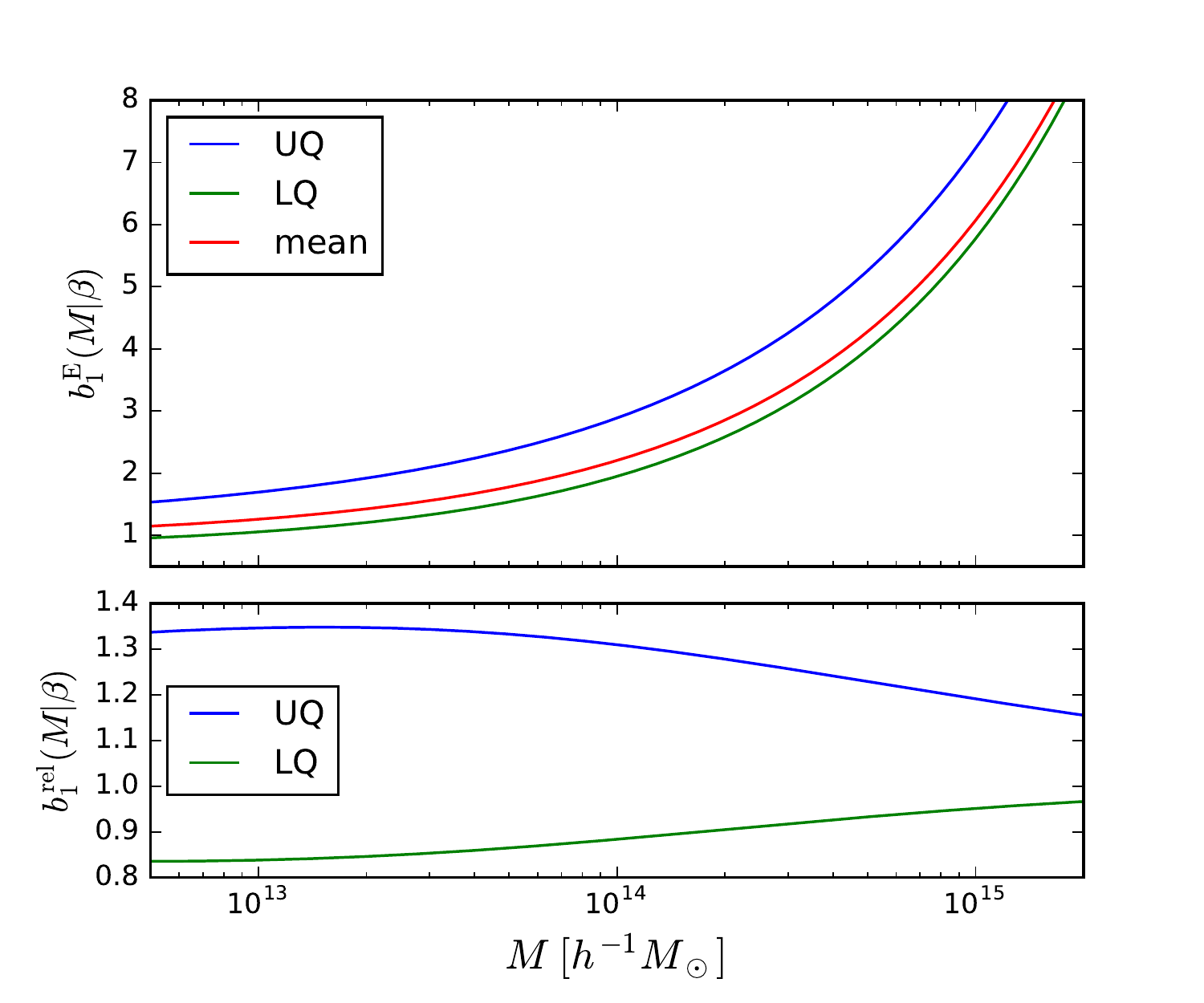}\
\caption{$\bie{1}$ and $\bir{1}$ as a function of halo mass, averaged over all values of the $\beta$ parameter (red curve), and over the upper and lower quartile in $\beta$ only (blue and green curve respectively).}
\label{fig:b1mass}
\end{figure}

Unfortunately, although $\beta$ carries a signal of assembly bias, it is not easy to relate it to halo properties that can be measured directly in simulations. 
In order to get some quantitative information we thus take a phenomenological approach and average only over the upper and lower quartiles of the distribution $p(\beta)$: this would quantify the amount of assembly bias obtained in terms of some final halo property that correlates perfectly with $\beta$. The results are shown in \reffig{b1mass}. As $\beta$ correlates with the shear strength, it would be natural to compare it with measurements of quantities describing the anisotropy of the final halo, like those presented in \cite{Faltenbacher:2009}, and in particular their anisotropy of the velocity dispersion. Comparison of the results shows the same qualitative trends and a good quantitative agreement overall. We note however that $\beta$ plays the opposite role of the various anisotropy parameters studied by \cite{Faltenbacher:2009}: while strongly biased protohalos with large $\beta$ correlate with highly sheared configurations with large ellipticity, the more strongly biased subsamples in \cite{Faltenbacher:2009} seem to correlate with more spherical final configurations. Nevertheless, this is only an apparent contradiction, because there is no model that relates $\beta$ (nor the initial amount of shear) to properties of the final halos. 

The second quantity that is known to give assembly bias in excursion sets with correlated steps (and therefore also in the ESP model) is the slope of the excursion set trajectory, which is strongly correlated with the peak curvature $x$. The observable property that can be naturally associated with $x$ is halo concentration. Large values of $x$ correspond to steep initial profiles and thus to halos that assemble most of their mass early, being therefore more concentrated. On the other hand, since a steeper slope implies a lower large-scale Lagrangian density, highly concentrated halos are expected to be less biased \cite{Dalal:2008,Musso:2014b}. This effect will be however slightly blurred in ESP by the simultaneous scatter in the barrier height.

Another halo property that can be easily accommodated within the excursion set framework is the mass accretion history \cite{Lacey:1993}. For excursion sets with correlated steps \cite{Musso:2012}, this leads to a statistically well-defined relation between the slope of the excursion set trajectories and the mass accretion rate. As each halo corresponds to one trajectory $\delta(\sigma)$, where $\delta$ varies as a function of the volume of the smoothing around the protohalo center, the mass assembly history can be inferred solving the implicit equation
\begin{equation}
\frac{\delta_c}{D(z)} = \delta[\sigma(M)] -\beta\sigma(M)\,,
\end{equation}
for $M(z)$ on the right-hand side.
Differentiating with respect to $z$ and then setting $z=0$ gives
\begin{equation}
  \frac{{\rm d}M}{{\rm d}z} = -\delta_c
  \bigg[\bigg(\frac{{\rm d} \delta}{{\rm d} \sigma}-\beta\bigg)\frac{{\rm d}\sigma}{{\rm d}M}\bigg]^{-1}
  \frac{{\rm d} D}{{\rm d}z}\,,
\end{equation}
at fixed $M$ and $\beta$, with ${\rm d}\delta/{\rm d} \sigma>\beta$ from the upcrossing condition. As the normalized slope correlates very strongly with the peak curvature $x$ (the two are actually identical for a Gaussian filter), this relation can be written  as
\begin{equation}
  \frac{\gamma\nu}{x-\gamma\beta} =
  \bigg|\frac{\mathrm{d}\log\sigma}{\mathrm{d}\log M}\bigg|
  \frac{1}{M}
  \frac{\mathrm{d}M/\mathrm{d}z}{\mathrm{d}D/\mathrm{d}z}
  \equiv\alpha\,,
\end{equation}
where $\alpha$ is proportional to the accretion rate.
We notice that, since $\mathrm{d}D/\mathrm{d}z<0$, the upcrossing condition $x>\gamma\beta$ restricts $\mathrm{d}M/\mathrm{d}z$ to negative values only, consistently with the fact that by construction excursion sets halos can only increase their mass. 

Inserting a Dirac delta $\delta_{\mathrm{D}}(x-\gamma\beta-\gamma\nu/\alpha)$ in \refeq{fESP} allows us to write the multiplicity function at fixed accretion rate as
\begin{equation}
\nu f(\nu,\alpha) = \frac{V}{V_*}
\frac{\nu e^{-\nu^2/2\Sigma^2}}{\sqrt{2\pi\alpha^2(1-\gamma^2)}}
\int {\rm d}\beta \,p(\beta)
F(\gamma\beta+\gamma\nu/\alpha)
\frac{e^{-(\nu+\beta)^2/2}}{\sqrt{2\pi}}\,,
\label{eq:fofnualpha}
\end{equation}
where $\Sigma^2\equiv \alpha^2(1-\gamma^2)/[\gamma^2(1-\alpha)^2]$, and the accretion rate $M^{-1}|{\rm d}M/{\rm d}z|$ only enters through $\alpha$ (whose residual dependence on $M$ is rather weak).
Crucially, in this expression the conditional distribution for $x$ given $\nu+\beta$ that appeared in \refeq{fESP} no longer depends on $\beta$ and comes out of the integral.
Differentiating this expression with respect to $\nu$, as described in \refeq{biastheor}, gives the Lagrangian bias coefficients $b^L_n(\nu,M^{-1}|{\rm d}M/{\rm d}z|)$ and allows one to evaluate their explicit dependence on the accretion rate. The first coefficient is
\begin{equation}
  b_1^L(\nu,\alpha) 
  = \frac{H_2(\nu/\Sigma)}{\nu\sigma} +
  \frac{\<[H_1(\nu+\beta)F-(\gamma/\alpha)F']\>_\nu}{\<F\>_\nu\sigma}\,,
\label{eq:ESPb1}
\end{equation}
where $F'(\gamma\beta+\gamma\nu/\alpha)$ is the derivative of $F$ with respect to its argument, $H_n$ is the $n$-th order Hermite polynomial, and we used the notation $\<\dots\>_\nu\equiv\int {\rm d}\beta\dots p(\beta)e^{-(\nu+\beta)^2/2}/\sqrt{2\pi}$.
Since $\Sigma\propto\alpha$, the first term in \refeq{ESPb1} grows as $1/\alpha^2$ in the limit of small accretion rate. Conversely, since $F(x)\sim x^3$ for large values of its argument, the second term remains finite. Thus, the linear bias scales as 
\begin{equation}
  b_1^E(\nu,\alpha) 
  \sim \frac{\Gamma^2}{\alpha^2\sigma}\nu\,,
\label{eq:b1scaling}
\end{equation}
in the small-$\alpha$ limit. Furthermore, as this term does not contain $p(\beta)$, we expect this effect to be rather model independent.

\begin{figure}
\centering
\includegraphics[scale=0.5]{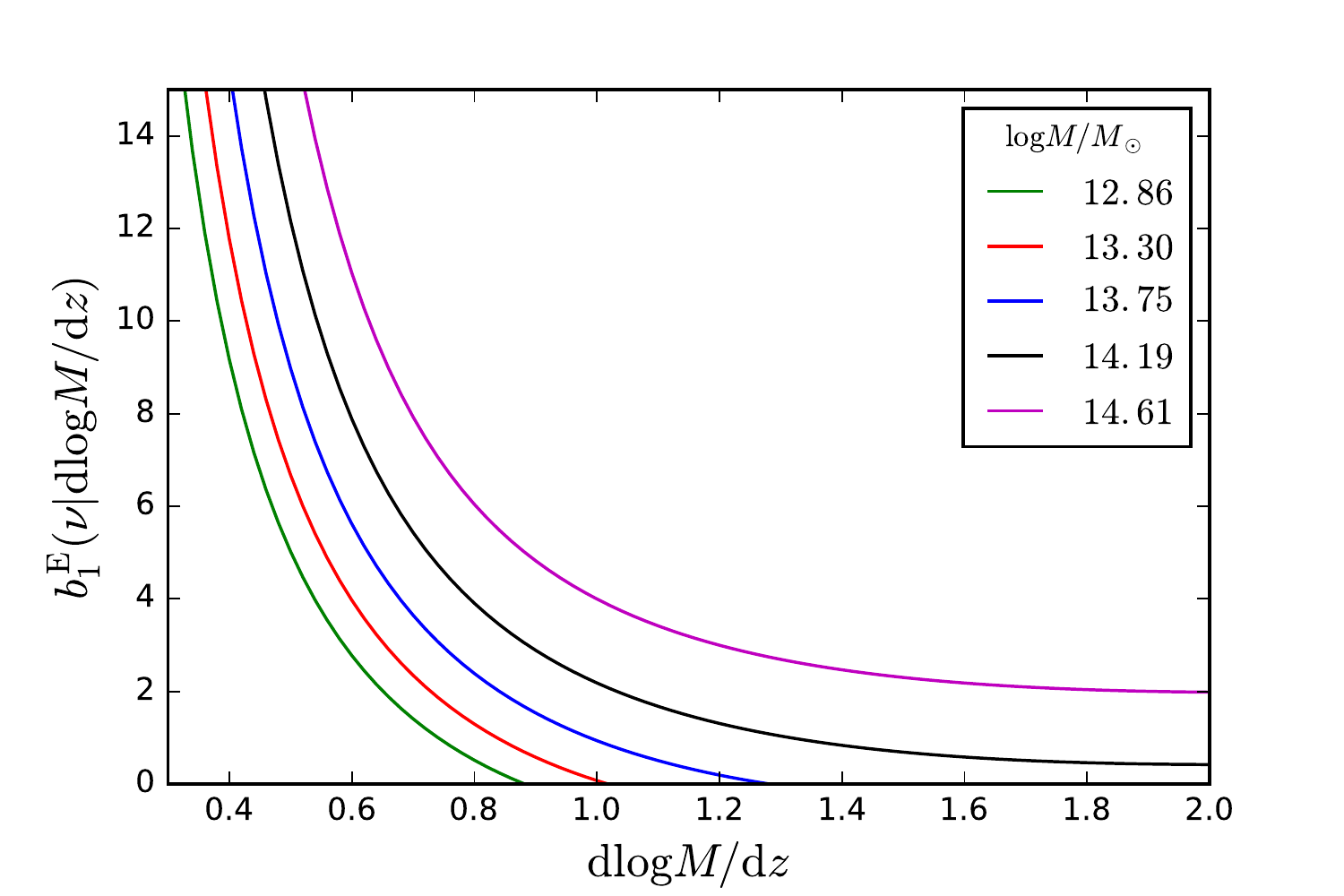}
\includegraphics[scale=0.5]{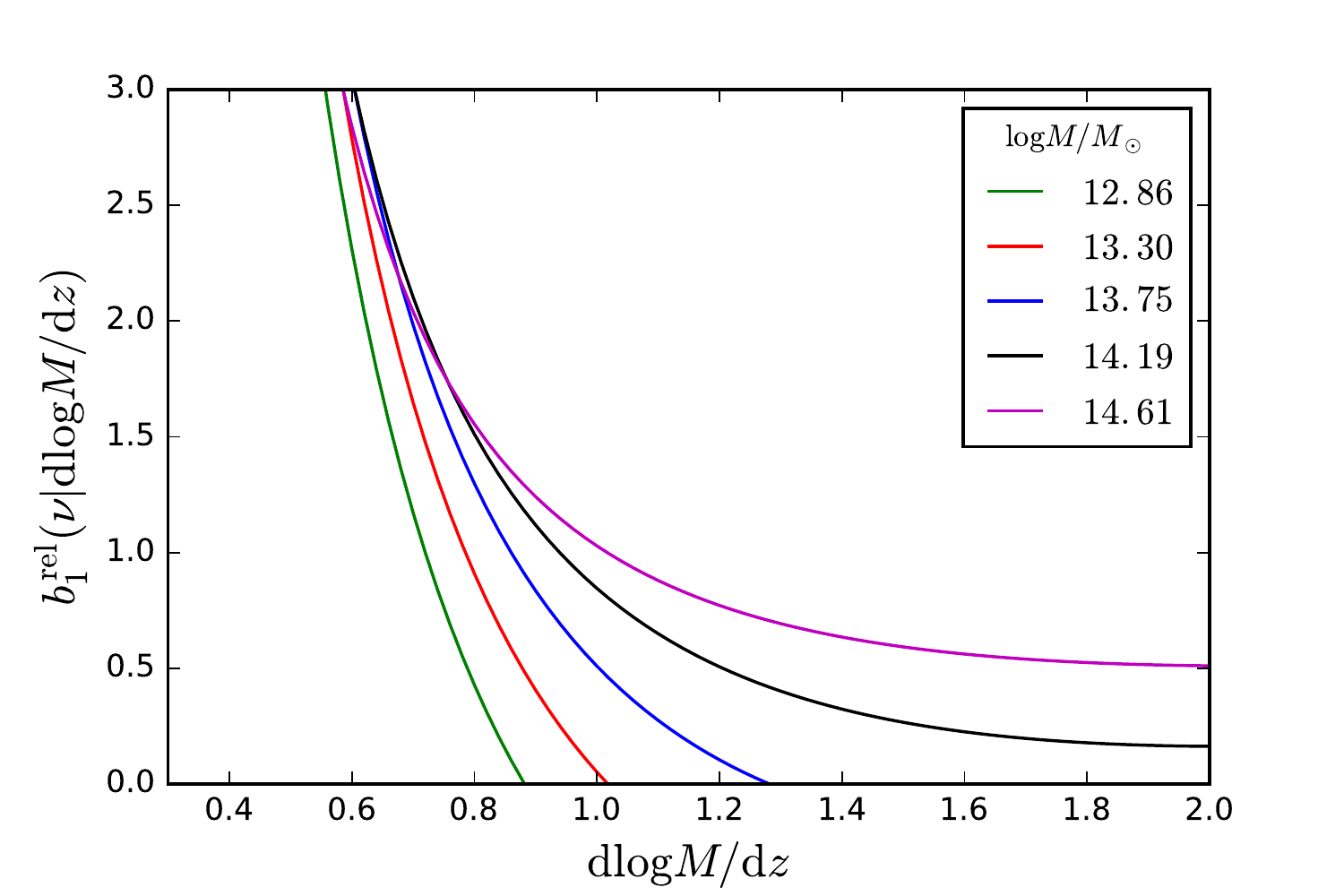}
\caption{$\bie{1}$ and $\bir{1}$ as a function of the theoretical mass accretion rate $M^{-1}|\mathrm{d} M/\mathrm{d} z|$ for several halo masses. }
\label{fig:b1Mdot}
\end{figure}

The full results are displayed in \reffig{b1Mdot}, showing that, at fixed mass, halo bias is indeed a decreasing function of accretion rate. Furthermore, the scaling \refeq{b1scaling} of the bias in the low-accretion-rate limit is a rather general feature that does not depend on the form of $p(\beta)$ nor of the barrier. While accretion rate is often thought to play the same role as concentration, we see that this is actually not the case.
A more accurate interpretation is then that slow accretion means that $x-\gamma\beta=\gamma\nu/\alpha$ must be much higher than, but proportional to, its expected value $\gamma\nu$: these halos are unlikely to climb such a steep density gradient relative to the barrier, and tend to be in the exponential tail of the conditional mass function. As such, their relative abundance is thus significantly more sensitive to a small change of the large-scale density (just like high-mass halos in the tail of the unconditional mass function are). Steep slope (relative to the barrier) thus means large bias. The same qualitative effect, though much milder, should be expected for halos with very high accretion rates.

%%%%%%%%%%%%%%%%%%%%%%%%%%%%%%%%%%%%%%%%%%%%%%%%%%%%%%%%%%%%%%%%%%%%%%%%%%%%%
%%%%%%%%%%%%%%%%%%%%%%%%%%%%%%%%%%%%%%%%%%%%%%%%%%%%%%%%%%%%%%%%%%%%%%%%%%%%%
\section{Assembly bias from separate universe simulations}
\label{sec:bsep} 

\subsection{Separate universe simulations and halo finding}
\label{sec:SUandH}

In this section, we briefly describe the characteristics of our sets of simulations and the halo finding procedure.  We refer the reader to \cite{Lazeyras:2015,Wagner:2015} for more details.  

Our results are based on the suite of separate universe simulations described in \cite{Wagner:2014,Wagner:2015}, performed using the cosmological code GADGET-2 \citep{Springel:2005}. The idea of the separate universe simulations is that a uniform matter overdensity $\dr$ of a scale larger than the simulation box can be absorbed in the background density $\tilde{\rho}$ of a modified cosmology simulation (throughout the whole paper, quantities in modified cosmologies will be denoted with a tilde), where
\be 
\tilde{\rho}(t) = \rho(t)\left[1+\dr(t)\right], 
\label{eq:dr}
\ee
with $\rho$ the mean matter density in a simulation with no overdensity (which we call the fiducial cosmology). Indeed, a uniform density can only be included in this way, since the Poisson equation for the potential enforces a vanishing mean density perturbation over the entire box. Thus one can understand a simulation with a constant overdensity $\dr$ as a separate universe simulation with a properly modified cosmology. Qualitatively, a positive overdensity causes slower expansion and enhances the growth of  structure, i.e. more halos, whereas a negative one will have the opposite effect. The precise mapping of $\dr$ to modified cosmological parameters is described in \cite{Wagner:2014}.  Crucially, we work to fully nonlinear order in $\dr(t)$.  

We use three sets of simulations denoted by ``lowres'', ``midres'' and ``highres'' throughout the paper. The parameters of the three sets are summarized in \reftab{sims}.
\begin{table} 
\begin{center}
\begin{tabular}{cccc}
\hline \vspace{-0.3cm} \\
Name	& $N_p$		& L $[h^{-1} {\rm Mpc}]$	& Realisations \\
\hline \vspace{-0.3cm} \\
lowres	& $256^3$ 	& 500							& 64 \vspace{0.15cm}\\
midres 	& $512^3$ 	& 500							& 16 \vspace{0.15cm}\\ 
highres 	& $512^3$ 	& 250							& 8  \\  
\hline
\end{tabular}
\caption{Properties of the simulations sets. $N_p$ specifies the total number of particle and L the comoving box size in one dimension.}
\label{table:sims}
\end{center}
\end{table}
For all sets, we run the fiducial cosmology, i.e. $\dr=0$,  and simulations with values of $\dr$ corresponding to $\d_L$ = \{$\pm$0.5, $\pm$0.4, $+$0.35, $\pm$0.3, $+$0.25, $\pm$0.2, $+$0.15, $\pm$0.1, $\pm$0.07, $\pm$0.05, $\pm$0.02, $\pm$0.01\}, where $\d_L$ is the present-day linearly extrapolated matter density contrast.  
The comoving box size in the modified cosmology simulations is adjusted to match that in the fiducial cosmology. 
Hence, in the high redshift limit ($z\rightarrow \infty$ for which $\dr\rightarrow 0$) the physical size of the box is the same for all simulations whereas at the present time ($z=0$ in the fiducial cosmology) the physical size of the simulation box varies with $\dr$. However, this choice of the box size has the advantage that the physical mass resolution is the same within each set of simulations regardless of the simulated overdensity.  
The mass resolution is $m_p = 5.6\cdot 10^{11}h^{-1} M_\odot$ in the ``lowres'' set of simulations, $m_p=7\cdot 10^{10}h^{-1} M_\odot$ in the ``midres'' one and $m_p=8.8\cdot 10^{9}h^{-1} M_\odot$ in the ``highres'' one. Furthermore, for the ``lowres'' set of simulation, we ran 64 realizations of the entire set of $\d_L$ values. For the ``midres'' and ``highres'' ones we ran only 16 and 8, respectively, realizations of each $\d_L$ value as they are more costly in terms of computation time.  
Each simulation was initialized using 2LPT at $z_i$ = 49.  

The halos were identified using the Amiga Halo Finder (hereafter AHF) \cite{Gill:2004,Knollmann:2009}, which identifies halos with a spherical overdensity (SO) algorithm.  We identify halos at a fixed proper time corresponding to $z=0$ in the fiducial cosmology.  
In this paper, we only use distinct halos and do not consider their sub-halos.  Further, we will restrict to halos with at least 500 particles within $r_{200}$ (the radius corresponding to an overdensity of 200 with respect to the background density) to ensure convergence of the halo properties considered, such as concentration and ellipticity.  

The key point in identifying halos with the spherical overdensity criterion is the setting of the density threshold.  We choose here a value of $\Delta_{\rm SO}=200$ times the background matter density in the \emph{fiducial} cosmology.  
Thus, our measured bias parameters are valid for this specific halo definition.  
For the simulations with a different background density, the threshold must be rescaled in order to compare halos identified using the same physical density in each simulation. Specifically, we need to use
\begin{equation}
\Delta_{\rm SO} = \frac{200}{1+\dr}\,.
\label{eq:DSO}
\end{equation}

Finally, following \cite{Lazeyras:2015}, we do not remove unbound particles (i.e particles which are not gravitationally bound to the halo they are located in) from halos. As argued in there, this effect should be very small (of order 1\% on the mass function).

\subsection{Assembly bias}
\label{sec:assemblybias}

The assembly bias of dark matter halos is broadly defined as the dependence of the bias on any other halo property than its mass. To study this effect we must thus first count halos in mass bins. We choose top-hat mass bins given by 
\be
W_n(M,M_{{\rm center}})=\begin{cases} 1 &\mbox{if } \left|{\rm log}(M)-{\rm log}(M_{{\rm center}})\right| \leq 0.225 \\
0 & \mbox{otherwise, } \end{cases}
\label{eq:bins}
\ee
where $M$ is the mass ($M_{{\rm center}}$ corresponding to center of the bin) and $\log$ is the base 10 logarithm.
  
We count halos in five bins centred from ${\rm log}\left(M_{{\rm center}}\right) = 12.875$ to ${\rm log}\left(M_{{\rm center}}\right) = 14.675$. We choose this fairly wide binning centred around quite high mass values (the typical mass of collapsing object is given by $\log\left(M_\star\right)=12.415$ in our cosmology) to ensure keeping halos with a minimum of 500 particles (this is important in order for quantities such as the halo concentration to be well defined) and to have enough halos per mass bin to build robust statistics. The highres set of simulations covers the four lowest mass bins, the midres one the three bins centred from ${\rm log}\left(M_{{\rm center}}\right) = 13.775$ to ${\rm log}\left(M_{{\rm center}}\right) = 14.675$. We use only one mass bin centred around ${\rm log}\left(M_{{\rm center}}\right) = 14.675$ for the low resolution simulations. We perform a weighted average of the results of various sets of simulations for the mass bins covered by more than one set of simulations. Finally, we show results at the mean mass $\bar M$ of a given bin calculated as
\be
\bar{M}=\frac{\int_{\rm Bin}{ {\rm d}m \, {\rm d}n/{\rm d}m \, m}}{\int_{\rm Bin}{ {\rm d}m \, {\rm d}n/{\rm d}m}}
\label{eq:meanm}
\ee
where ${\rm d}n/{\rm d}m$ is the halo mass function.  Given the finite mass bin width, this can be numerically important when comparing results with previous work or theoretical predictions.

We now turn to other halo properties. As stated in the introduction, by further subdividing each mass bin with respect to one more quantity, we can evaluate the bias parameters at fixed mass as a function of any property (i.e. the assembly bias) with \refeq{asblybnsepuniv}. To do this, we closely follow the procedure outlined in \cite{Lazeyras:2015}, section 3.2. Specifically, we compute
\be
\delta_h (M,p,\d_L) = \frac{\tilde{N}(M,p,\d_L) - N(M,p)}{N(M,p)},
\label{eq:DN_N}
\ee
with $\tilde N(M,p,\d_L)$ the number of halos in a mass bin centred around mass $M$ and other property bin centred around $p$ in the presence of the linear overdensity $\d_L$ and $N(M,p)=\tilde N(M,p,\d_L=0)$. 
Note that $\d_h(M,p,\d_L)$ is the overdensity of halos in Lagrangian space as the separate universe simulations have the same comoving rather than physical volume.  

In order to obtain the Lagrangian bias parameters $b_n^L$, we then fit \refeq{DN_N} by  
\be
\d_h = \sum_{n=1}^5  \frac{1}{n!} b_n^L (\d_L)^n\,,
\label{eq:BiasExp}
\ee
i.e. we fit a fifth order polynomial to the relation $\d_h(\d_L)$. Ref. \cite{Lazeyras:2015} studied the effect of the degree of the polynomial on the results; as a rough rule, if one is interested in $b_n^L$, then one should fit a polynomial up to order $n+2$ (see their Appendix C). This procedure allows us to read off the Lagrangian bias parameters as the best fit coefficients directly. We then derive the Eulerian bias parameters $b^E_n$ from the measured Lagrangian ones, for which the fitting is more robust, using the exact nonlinear evolution of $\dr$ (\cite{mo/white:1996}; see Appendix B of \cite{Lazeyras:2015} for the details of the mapping). In order to estimate the overall best-fit of and error bars on the bias parameters, we follow the bootstrap technique outlined in \cite{Lazeyras:2015}.

We investigate the dependence of the halo bias on four halo properties: the concentration $c_V$, spin parameter $\lambda$, logarithmic mass accretion rate $M^{-1}{\rm d}M/{\rm d}z$ and shape $s$. The halo concentration is quantified using the usual NFW concentration parameter $c_V$ measured as in \cite{Prada:2011}. More specifically, AHF computes the ratio between the maximum of the circular velocity $V_{\rm max}$ and $V_{200}$, the circular velocity at $r_{200}$. For the case of the NFW halo profile, this ratio is given by
\be
\frac{V_{\rm max}}{V_{200}}=\left(\frac{0.216 \, c_V}{f(c_V)}\right)^{1/2},
\label{eq:cprada}
\ee
where $f(c_V)$ is given by
\be
f(c_V)=\ln (1+c_V)-\frac{c_V}{1+c_V}.
\label{eq:fofc}
\ee
Computing $c_V$ from the circular velocity at two different radii is hence straightforward. However, as we will see, this way of inferring the concentration is not as robust as a proper fit of the halo density profile. For the halo spin, we use the spin parameter as defined in \cite{Bullock:2000} 
\be
\lambda = \frac{|\v{J}|}{\sqrt{2}MVr_{200}},
\label{eq:spinparam}
\ee
where the angular momentum $\v{J}$, the mass $M$ and the circular velocity $V$ are evaluated at position $r_{200}$. 
Using the AHF MergerTree code, we also compute the mass accretion rate between $z=0.5$ and $z=0$ normalized to the final halo mass 
\be
\frac{1}{M}\frac{{\rm d}M}{{\rm d}z} \equiv \frac{1}{M(z=0)}\frac{M(z=0)-M(z=0.5)}{|\Delta z|},
\label{eq:mdot}
\ee 
for $|\Delta z|=0.5$.  We choose this redshift interval to ensure that we have a corresponding time interval greater than the dynamical time of a halo. In addition, we allow this quantity to be negative by at maximum -1. This is to avoid considering extremely stripped low-mass halos in the vicinity of a massive halo.  As shown in \refapp{massev}, \reffig{Mdothist}, this only removes a very small fraction of halos and should not affect the results.  
Finally, following the work of \cite{Faltenbacher:2009}, we also measure the bias as function of halo shape given by
\be
s=\frac{c}{a},
\label{eq:shape}
\ee 
where $a > b > c$ are the axes of the moment-of-inertia tensor of the halo particles.

For each mass bin, we divide halos into four bins for each of these quantities, determined as the four quartiles of the distribution at fixed mass in the fiducial cosmology (i.e using all 64, 16 or 8 realisations). We determine the bins using the set of simulations providing the most volume at the given mass bin. 
For the halo concentration, spin and shape, we follow W06 and define
\be
p'= \frac{{\rm ln}(p/\bar{p})}{\sigma({\rm ln}p)},
\label{eq:prime}
\ee 
where $p$ is the mean of $c_V$, $\lambda$ or $s$ in a given quartile, $\bar{p}$ is the mean in a given mass bin and $\sigma$ is the square root of the variance at fixed mass. 
The logarithm is not defined for the logarithmic mass accretion rate as we allow for it to be negative or zero. We hence only compute the difference from the mean of this quantity in each quartile.

%%%%%%%%%%%%%%%%%%%%%%%%%%%%%%%%%%%%%%%%%%%%%%%%%%%%%%%%%%%%%%%%%%%%%%%%%%%%%
%%%%%%%%%%%%%%%%%%%%%%%%%%%%%%%%%%%%%%%%%%%%%%%%%%%%%%%%%%%%%%%%%%%%%%%%%%%%%
\section{Results and discussion}
\label{sec:results}

We now turn to our results. \refSec{abiasres} presents the assembly bias in $b^E_1$ and $b^E_2$ as a function  of the four properties presented before and discusses these results. In \refsec{2bin}, we present assembly bias in $b^E_1$ with respect to two halo properties. Finally, in order to lighten the notation, we drop the mass argument in the bias parameters in the following when unnecessary. It should be understood that all the relations we describe are at fixed mass.

\subsection{Assembly bias as a function of halo concentration, spin, mass accretion rate and shape}
\label{sec:abiasres}

\begin{figure}[t]
\centering
\includegraphics[scale=0.37]{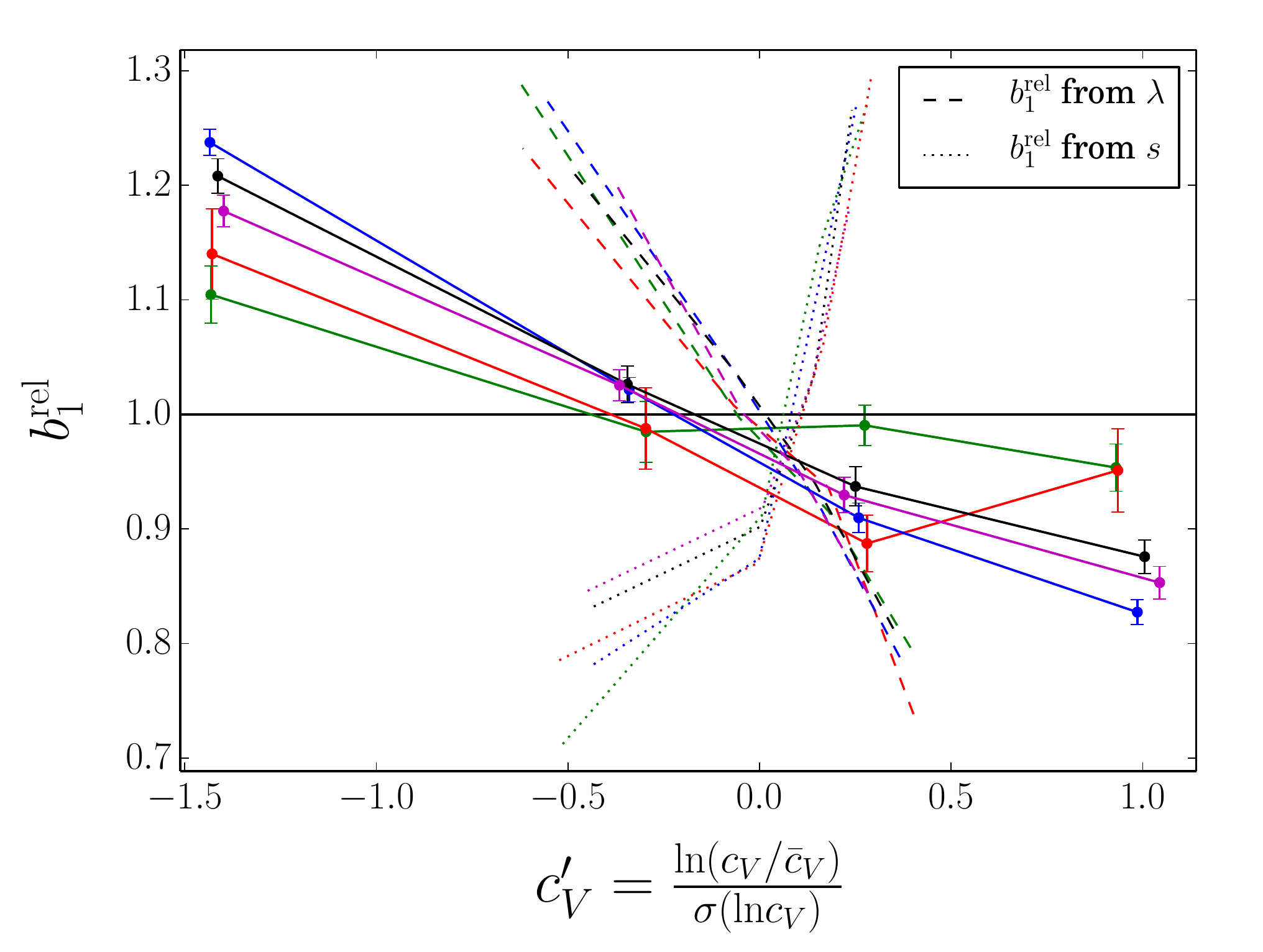}
\includegraphics[scale=0.37]{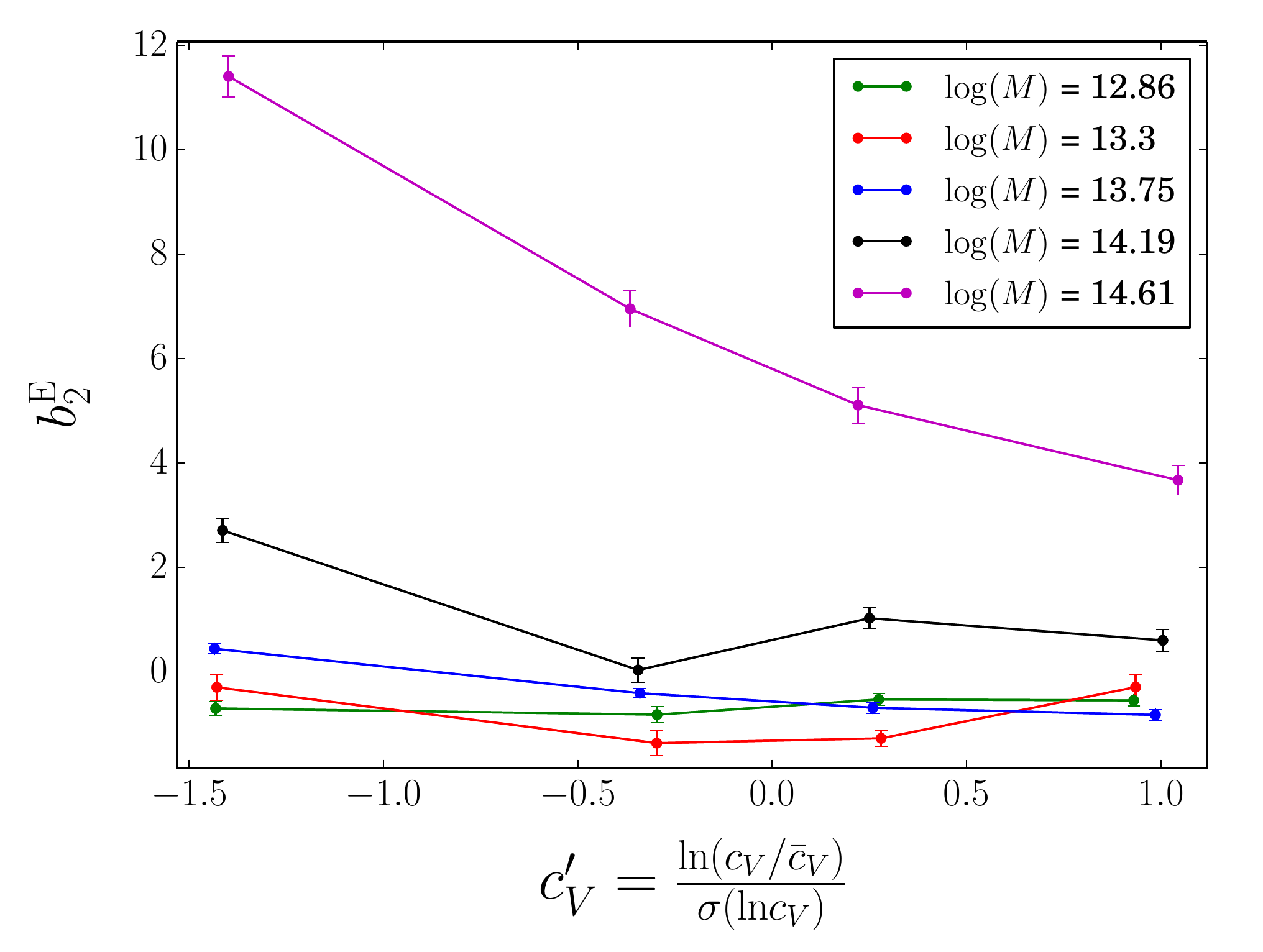}
\caption{\textbf{Left panel:} Linear assembly bias $\bir{1}$ as a function of concentration for several mass bins (indicated by the color code). The points linked by solid lines show our direct measurements from separate universe simulations. The errorbars show the $1\sigma$ bootstrap error. The dashed and dotted curves show the reconstruction of these relations using \reffig{bofL} and \reffig{bofs} (respectively) and the mean relations presented in \reffig{cofL} and \reffig{cofs} (respectively) in \refapp{haloprop}. \textbf{Right panel:} Eulerian quadratic bias as a function of concentration as measured in separate universe simulations for the same mass bins.}
\label{fig:bofc}
\end{figure}

\refFigs{bofc}{bofMdot} present results for $b^E_1$ and $b^E_2$ as a function of concentration $c_V$, spin parameter $\lambda$, shape $s$, and mass accretion rate $M^{-1}{\rm d} M/{\rm d}z$, for our halo mass bins. The points linked by solid lines on the left panels of these figures show the relative linear bias $\bir{1}=\bie{1}\left(p |M \right)/\bie{1}(M)$ (where $\bie{1}(M)$ is the Eulerian linear density bias coefficient as measured in \cite{Lazeyras:2015}). In addition, we present the reconstructed assembly bias with respect to property $p_1$ using the assembly bias as a function of another property $p_2$ and the mean relation $p_1(p_2)$ (these relations are presented in \refapp{haloprop}). Each time, we present curves for the best (dashed curves) and worst (dotted curves)  reconstruction. As can be expected from previous works and as we will see, this reconstruction works very poorly in most cases. The right panels of \reffigs{bofc}{bofMdot} show our measurements for $\bie{2}$. The color coding, indicating the mass, is the same on each figure and for each set of curves.

Before getting into the detailed analysis of these figures, we would like to stress that, although we will refer to low and high masses, one should keep in mind that all our results are for masses above $M_\star$, and hence technically massive halos. 
Finally, as explained in \refsec{SUandH}, we used a spherical overdensity algorithm to identify halos. This most probably has an impact on our findings, but we do not investigate how they change if we used, e.g., a friends-of-friends (FoF) algorithm.

Throughout this section, we will quantitatively compare our results for the linear bias with previous results from \cite{Gao:2006, Faltenbacher:2009, Wechsler:2005}. There are not many previous results for assembly bias in higher order biases to compare with (see however \citep{Angulo:2007, Paranjape:2016}), and we therefore do not conduct any quantitative comparison for $\bie{2}$. When making the comparison for $b^E_1$ one should keep in mind various differences in the way the analysis was conducted in this work and in the previous ones. 

The first and dominant difference are the scales on which the bias is measured.  The three aforementioned previous studies have estimated the halo bias through the real space 2-point correlation function on scales much smaller than the ones considered in our work;  Refs.~\cite{Gao:2006, Faltenbacher:2009} use comoving scales from $6-20$ Mpc, while \cite{Wechsler:2005} use scales in the range $5-10\Mpch$. On such small scales, nonlinear effects are relevant, so that their inferred linear bias $b^E_1$ does not directly correspond to the (renormalized) bias parameter in the large-scale limit which we measure here.  

Secondly, \cite{Gao:2006, Faltenbacher:2009} use an FoF halo finder (\cite{Davis:1985}) while \cite{Wechsler:2005} use a variant of the bound maxima algorithm (\cite{Klypin:1999}). This could also have an important effect as it implies that we do not study exactly the same objects.  A final, though likely subdominant, difference is the use of different background cosmologies in these studies.

\begin{figure}[t]
\centering
\includegraphics[scale=0.37]{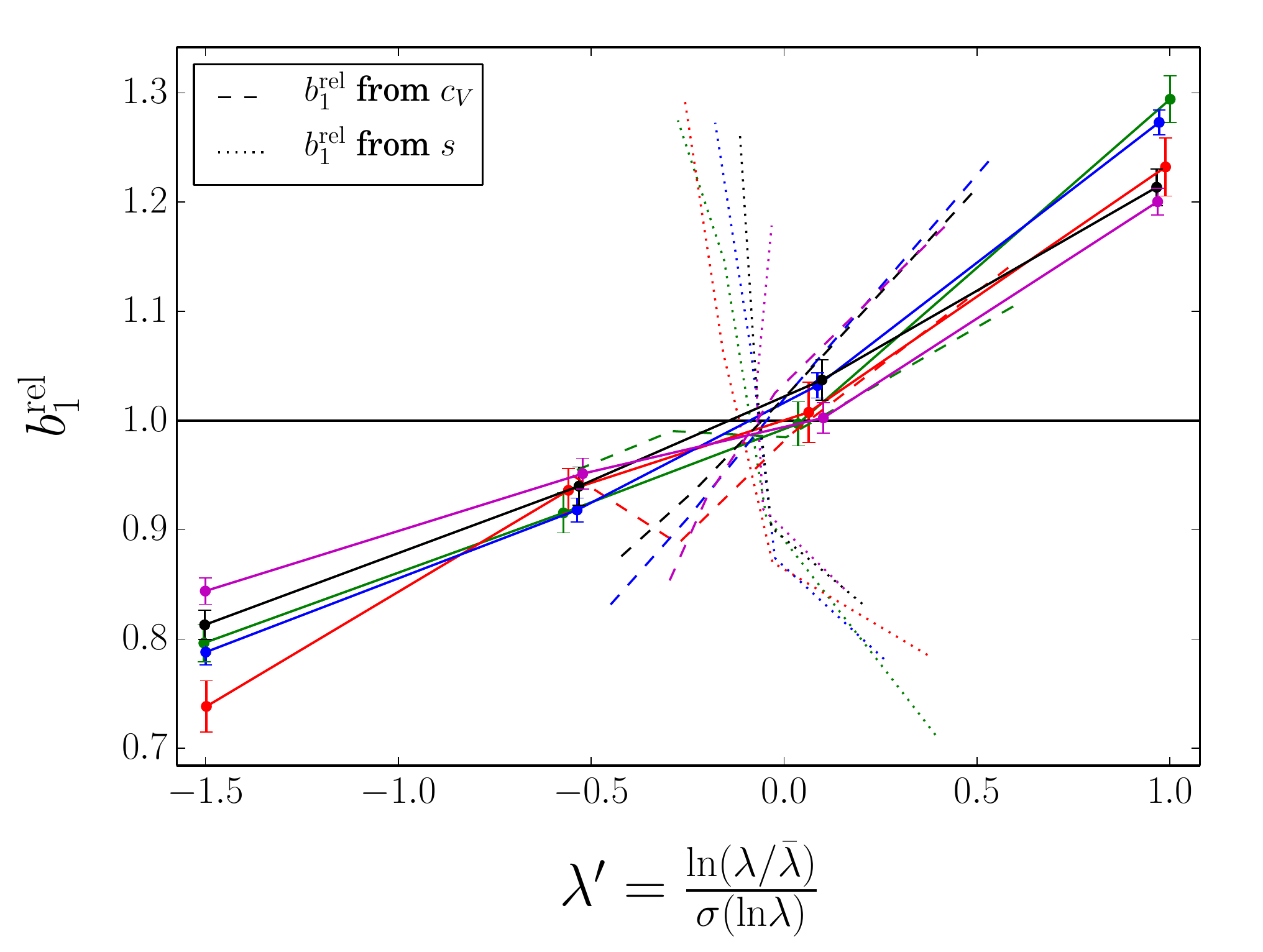}
\includegraphics[scale=0.37]{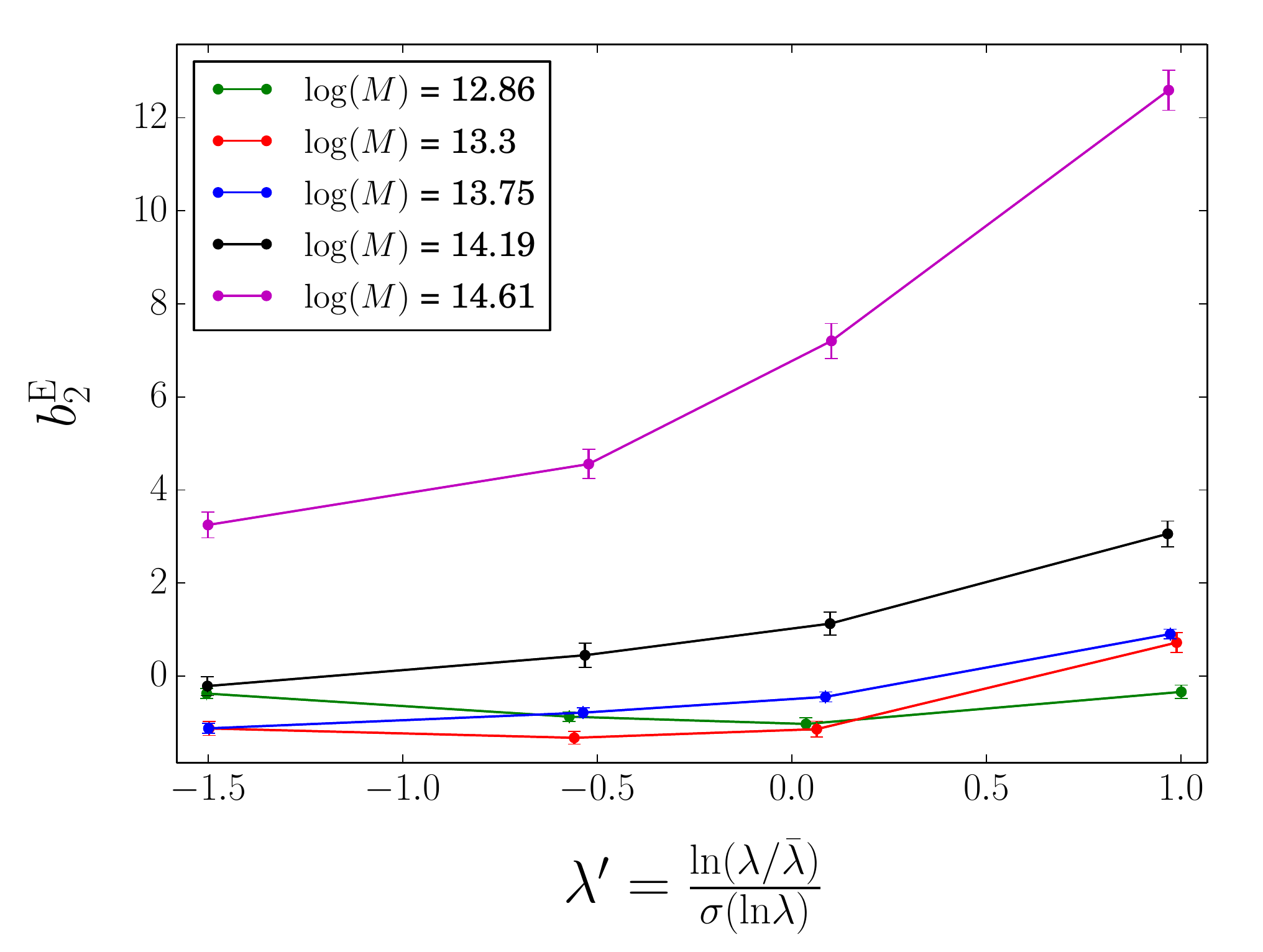}
\caption{Same as \reffig{bofc} but for $\bir{1}$ and $\bie{2}$ as a function of the spin parameter $\lambda$.}
\label{fig:bofL}
\end{figure}

The results for $\bir{1}$ as a function of concentration (left panel of \reffig{bofc}) are in qualitative agreement with previous studies (e.g. \cite{Gao:2006,Wechsler:2005}), i.e we find that for mass $M>M_\star$, more concentrated halos are less clustered. This effect is strongest at intermediate mass (i.e around $\log (M)=13.75$) and decreases monotonically for both higher and lower mass values. 
In \refapp{wechsler} we quantitatively compare our results to the best fit of W06 (see \reffig{bofcwechs}). All measurements are within a $1\sigma$ error region roughly inferred from their figure 4 (note however that our $1\sigma$ error bars are much smaller than theirs). One thing to be noticed on \reffig{bofcwechs} is that all our curves are convex, whereas W06's curves become concave at high mass. This effect is most probably due to the fact that we use the definition \refeq{cprada} for the concentration $c_V$ in our work, which is known to be a poor proxy for the concentration obtained from a full profile fit as in W06, especially at high masses.  Considering this as well as the differences pointed out above, the overall agreement of our results with theirs is very satisfying.

\begin{figure}[t]
\centering
\includegraphics[scale=0.37]{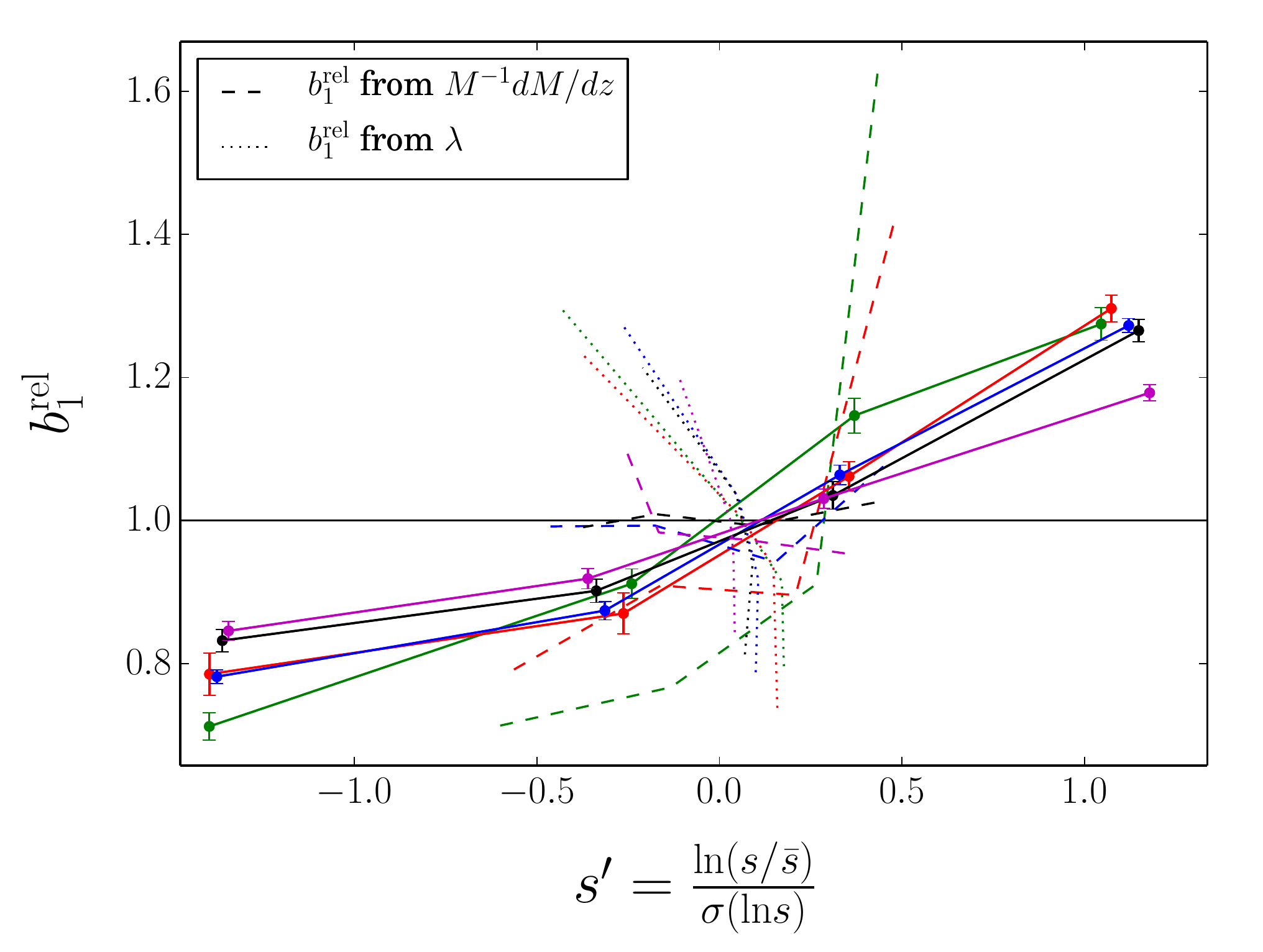}
\includegraphics[scale=0.37]{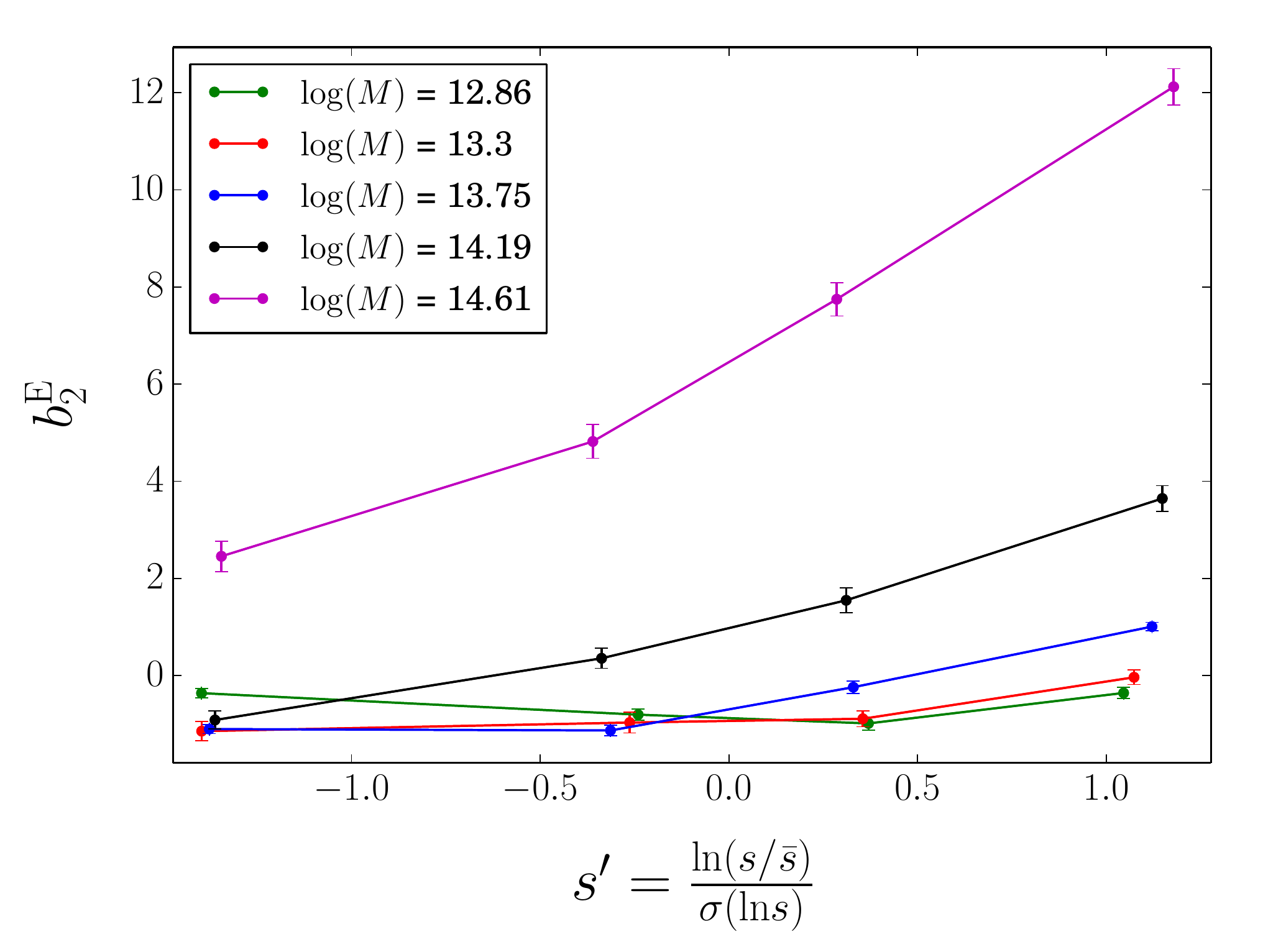}
\caption{Same as \reffig{bofc} but for $\bir{1}$ and $\bie{2}$ as a function of  the halo shape parametrized by $s=c/a$ the ratio of the smallest half axis of the inertia tensor to the largest one.}
\label{fig:bofs}
\end{figure}

We also inferred the assembly bias with respect to concentration using the assembly bias with respect to spin parameter and shape (\reffig{bofL} and \reffig{bofs}) and the mean relations between $c_V$ and these quantities presented in \reffigs{cofL}{cofs} in \refapp{haloprop}. The reconstruction using the spin $\lambda$ gives the best results while the one using the shape $s$ gives the worst (the reconstruction using the mass accretion rate, not presented here for sake of clarity, lies somewhere in between the two). Even though we did not expect this kind of reconstruction to work, it is interesting to see to what degree they disagree with the direct measurements. In the case of the shape, it is already clear from \reffig{bofs} and \reffig{cofs} that no good results could be obtained since the assembly bias as a function of $c_V$ and $s$ present opposite behaviours but the $c_V(s)$ relation shows a monotonic increase of $c_V$ with $s$: more spherical halos (i.e. with positive $s'$) have a higher concentration. 

The results for $\bie{2}$ as a function of concentration are presented in the right panel of \reffig{bofc}. We obtain a clear detection of assembly bias (especially at high mass) following the same trend as for $b^E_1$ (this confirms the recent findings of \cite{Paranjape:2016}, see their figure 10). The fact that assembly bias in the nonlinear bias parameter $b^E_2$ follows the same trend as that in $b^E_1$ could explain why our results agree with measurements from much smaller scales, which, in the perturbative framework, measure a combination of the large-scale bias parameters $b^E_1$ and $b^E_2$ as well as other higher-order bias parameters.

Turning to $\bir{1}$ as a function of spin (\reffig{bofL}), one sees that halos with more angular momentum are also more clustered. It is interesting to see that the relation $\bir{1}(\lambda)$ seems to be slightly less mass dependent than that in the concentration.  This is also in reasonable quantitative agreement with previous results \cite{Gao:2006}, which is not self-evident given the various differences in the way the respective analyses were conducted. The reconstructed curves show that the reconstruction works best when using the concentration relations (\reffig{bofc} and \reffig {Lofc}). These relations can actually almost reconstruct $\bir{1}(\lambda)$ correctly. This reflects the fact that high spin parameter halos have a particle distribution extending further from the halo center (leading to lower concentration), an effect that can be understood when considering the particle kinetic versus potential energy. On the other hand, the shape relations again work the worst at reconstructing the relative bias as a function of spin. Clearly then, the assembly bias of massive halos cannot be controlled by a single parameter beyond the halo mass. While the relatively higher-biased population of halos with high spin is probably roughly the same as that with low concentration, these halos do not correspond to those higher-biased halos that are more spherical than average. The same argument applies to the relatively less-biased populations, and is further supported by the fact that $c_V$ depends more strongly on $\lambda$ than $s$ (see \reffigs{cofL} {cofs}) and that the same behaviour can be observed for $\lambda$ as a function of $c_V$ and $s$ (\reffigs{Lofc}{Lofs}).  
The results for $\bie{2}$ again show that assembly bias is also present in this parameter. As for $\bie{2}$ as a function of concentration, the effect goes in the same way as for $b^E_1$ and seems more important at high mass.

\begin{figure}[t]
\centering
\includegraphics[scale=0.37]{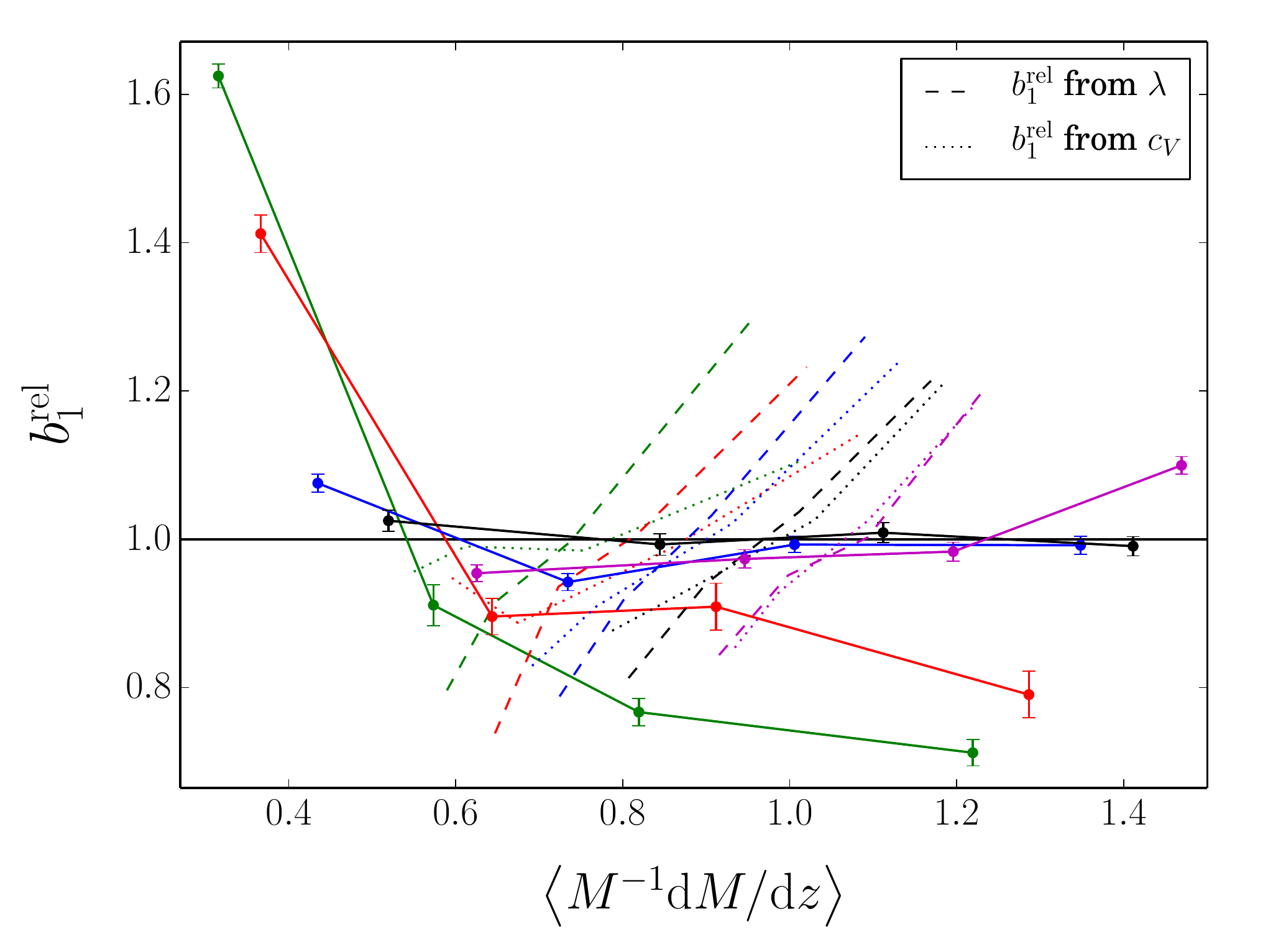}
\includegraphics[scale=0.37]{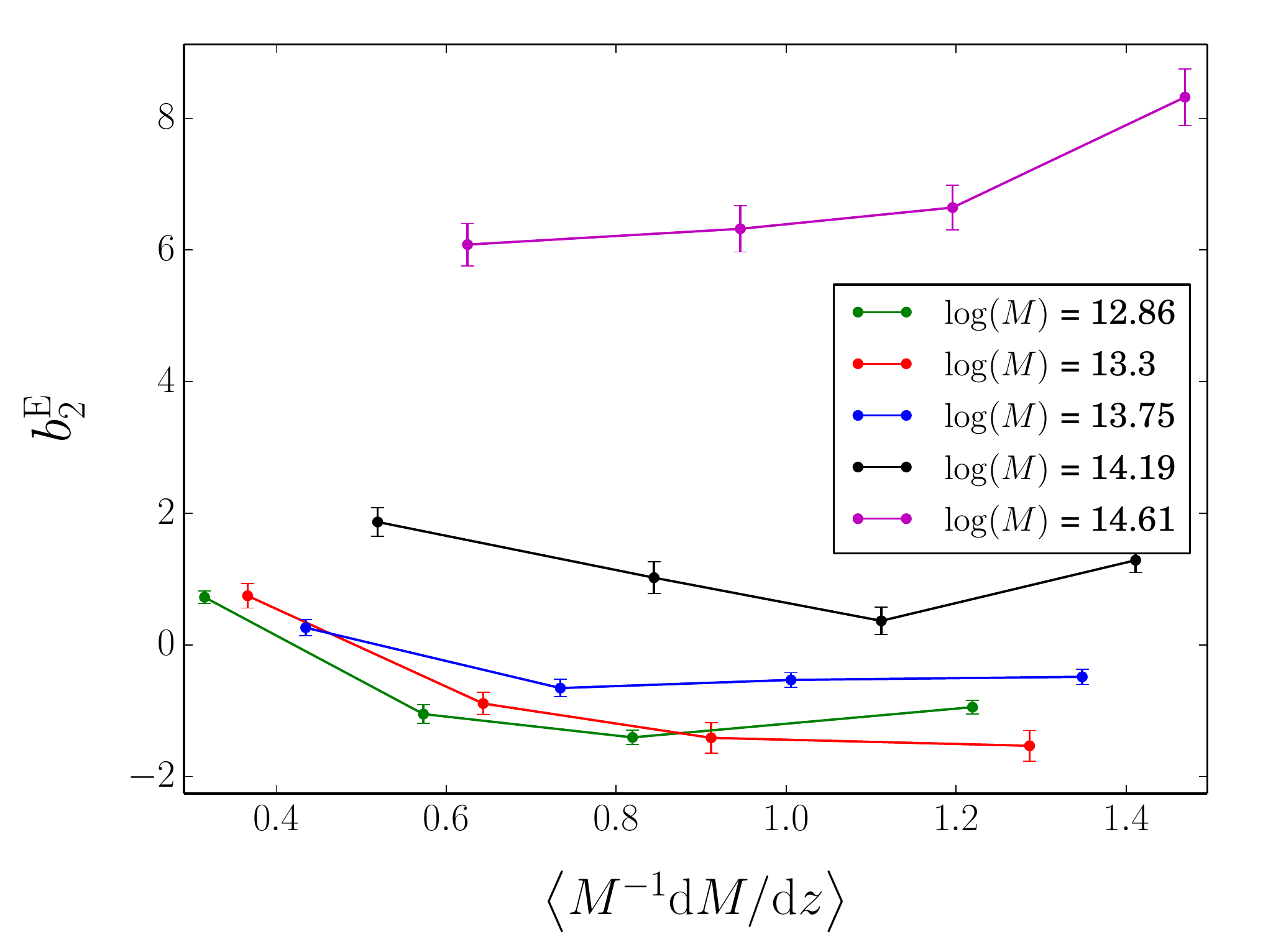}
\caption{Same as \reffig{bofc} but for $\bir{1}$ and $\bie{2}$ as a function of mass accretion rate.}
\label{fig:bofMdot}
\end{figure}

The assembly bias with respect to halo shape is shown in \reffig{bofs}. Once again we clearly detect assembly bias in both $b^E_1$ and $b^E_2$. More spherical halos (i.e. with positive $s'$) are more clustered. For $\bir{1}$ this behaviour is milder for higher mass halos. These results are in good quantitative agreement with the results from \cite{Faltenbacher:2009} (see their figure 1, top left panel).  
The reconstructed curve for $\bir{1}$ from the mass accretion rate is in better agreement than the one from the spin, which could be expected from the fact that the shape works very poorly to reconstruct the bias as a function of spin or concentration.  

Finally, \reffig{bofMdot} shows the dependence of $b^{\rm rel}_1$ and $b^E_2$ on the mass accretion rate. To our knowledge, this is first time that assembly bias with respect to this quantity is measured.   The effect is mass dependent and seems to reverse around $\log (M)\approx 14$ in both $\bir{1}$ and $\bie{2}$. Indeed, at high mass, halos accreting faster are more clustered whereas at low mass the opposite effect is observed. This result is quite unexpected, especially at low mass. Indeed, \reffigs{MdotofL}{Mdotofc} show that halos with a higher mass accretion are less concentrated and have higher angular momentum. Hence one would expect fast-accreting halos to be more clustered (as is shown by the reconstructed curves for $\bir{1}$, which all perform very poorly in this case). Note that the high bias at low mass accretion rate and mass could be partly due to tidally stripped halos (with negative mass accretion) in the vicinity of a much larger structure. As their distribution by definition follows that of massive halos, they are highly biased and could contaminate our low-mass-accretion-rate population, enhancing the increase of $\bir{1}$ and $\bie{2}$. We discuss this effect in more detail in \refapp{massev}. 

Ref.~\cite{Dalal:2008} measured the logarithmic derivative of the Lagrangian overdensity of halos with respect to smoothing scale, ${\rm d}\ln \d(M)/{\rm d}\ln M$, which, as discussed in \refsec{theory}, is the excursion set proxy for the inverse of the mass accretion rate. It is difficult to compare our results in \reffig{bofMdot} to their results, due to the nontrivial link of the Lagrangian slope ${\rm d}\ln\d(M)/{\rm d}\ln M$ with the late-time mass accretion rate estimated here. It seems however that the results for the bias as a function of the Lagrangian slope are more closely related to those as a function of the concentration. Indeed, a shallower slope corresponds to a lower concentration and both yield a higher bias at high mass (see figure 10, middle panel of \cite{Dalal:2008} and our \reffig{bofc}). The relation between the Lagrangian slope and the late-time mass accretion rate is more complicated. In the context of ESP (or any other model based on excursion sets), this can be seen from the fact that one needs to consider the \textit{logarithmic} slope as a proxy of mass accretion rate, which constrains the ratio of $\nu$ and slope, rather than the slope alone. This introduces a nontrivial dependence of $f(\nu,\alpha)$ on $\nu$ [\refeq{fofnualpha}] and prevents us from identifying the population of halos with a shallow Lagragian slope with that of halos having a higher late-time mass accretion rate. 
We can also compare our measurements with the theoretical prediction of the ESP on the right panel of \refFig{b1Mdot}. The ESP predicts a strong decrease of $\bir{1}$ at low mass accretion rates with a plateau toward higher $M^{-1} {\rm d}M/{\rm d}z$ values at high mass. This behaviour, that we expect to be rather model independent, is qualitatively what we see in our results, albeit only for low halo mass. The quantitative agreement is however rather poor. This is likely to be a consequence of the many approximations occurring when identifying, on a halo by halo basis, Lagrangian quantities in theoretical models like ESP with final halo properties measured in simulations.

Using the interpretation of the ESP stochastic variable $\beta$ as the large-scale shear field around halos we can try to link our results with what was presented on the left panel of \reffig{b1beta}. As already explained, in this picture, higher values of $\beta$ are interpreted as higher shear implying a higher halo ellipticity \textit{in Lagrangian space, i.e. in the initial conditions}. The way to relate $\beta$ to any late-time quantities is not established yet. We can however see that $\beta$ seems to behave as the inverse of the shape factor $s$ that we measured, as more spherical halos have a higher bias value, which is also the case for higher $\beta$ values. Results for assembly bias with respect to several anisotropy parameters were presented in \cite{Faltenbacher:2009} and, as previously stated, the behaviour in $\beta$ predicted from the ESP is inverse to all these results. If our interpretation of linking $\beta$ to the initial halo shape is correct, this would mean that, at all masses considered here, more elliptical halos (than average) in the initial conditions tend to end up as more spherical than average at final time.  This is unexpected and clearly warrants further investigation. 

Regarding $b^E_3$, our simulations do not allow for a clean detection of assembly bias, although evidence is seen for an assembly bias in $b^E_3$ with respect to each halo property.  In contrast to the results for the quadratic bias, there are indications that the effect does not always go in the same direction as for $b^E_1$, although this result is not highly significant statistically.

\subsection{Assembly bias with respect to two halo properties}
\label{sec:2bin}

\begin{figure}
\centering
\includegraphics[scale=0.37]{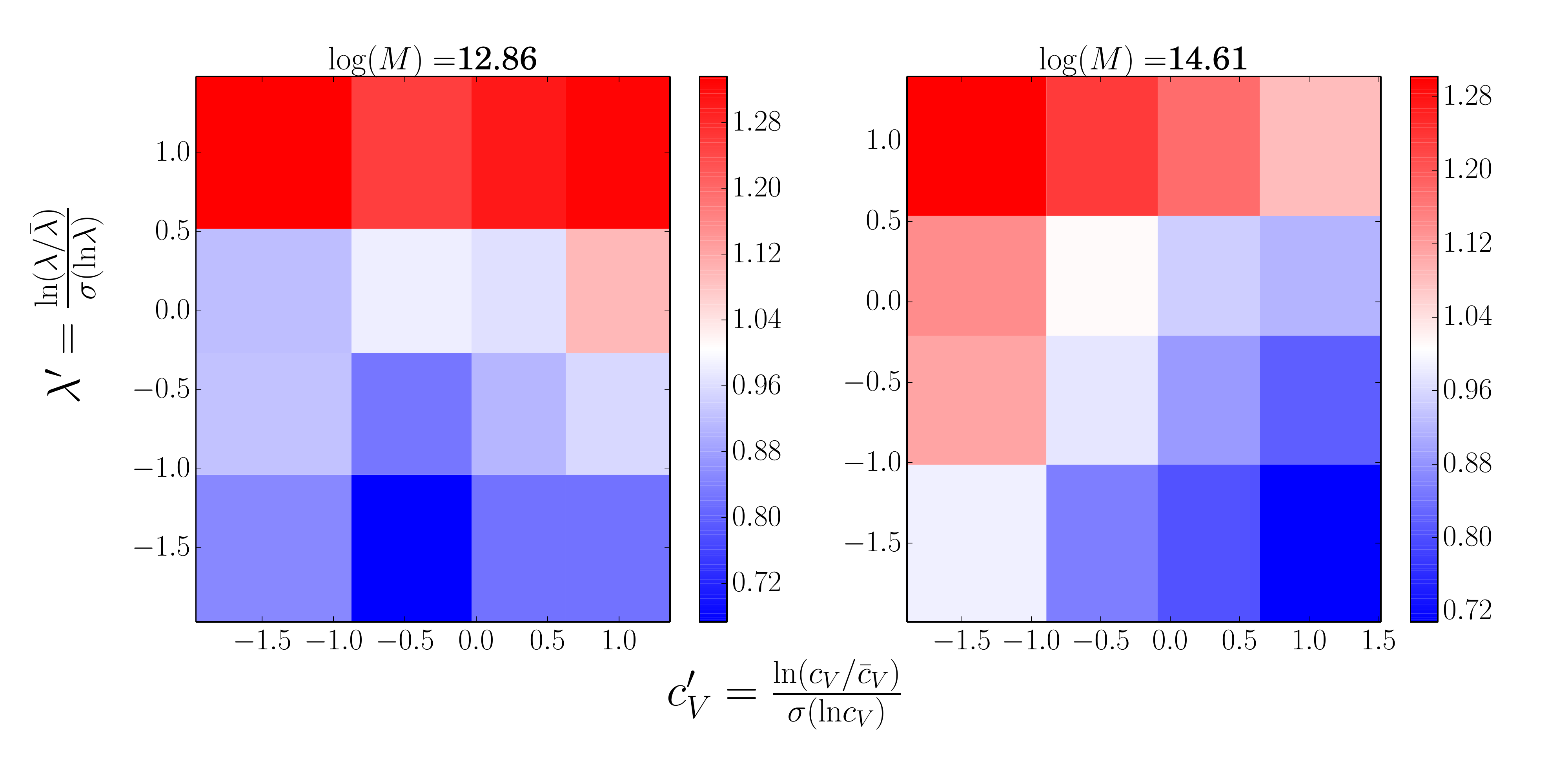}
\caption{$b_1^{\rm rel}$ as a function of halo spin parameter and concentration for the highest and lowest mass bins. A white cell corresponds to $b_1^{\rm rel}=1$.}
\label{fig:bofLandc}
\end{figure}

This section presents results for the linear relative bias parameter as a function of two halo properties. We focus on $\bir{1}(c, {\rm d}M/{\rm d}z)$, $\bir{1}(c, \lambda)$, $\bir{1}(c, s)$ and $\bir{1}(s, \lambda)$. The results are presented in \reffigs{bofLandc}{bofLands}. The $x$ and $y$ axes represent the two halo properties, while the color coding shows the amplitude of $\bir{1}(p_1,p_2)$ with red bins corresponding to higher relative bias and blue bins to lower bias. Each time we only show the lowest and highest mass bins. We verified that the evolution of $\bir{1}(p_1, p_2)$ with the halo mass in each bin is essentially monotonic which makes the presentation of results at intermediate mass unnecessary. The main idea when looking at these plots is to see whether the assembly bias as a function of one property changes when another halo property is also specified. We stress that the procedure to obtain these plots is the same as to obtain \reffigs{bofc}{bofMdot} except that after binning in $p_1$ we further compute quartiles of the distribution of the property $p_2$ conditioned on $p_1$ and $M$. We can then compute the assembly bias using \refeq{DN_N}, where the halo number now depends on two halo properties in addition to the mass. 

\begin{figure}
\centering
\includegraphics[scale=0.39]{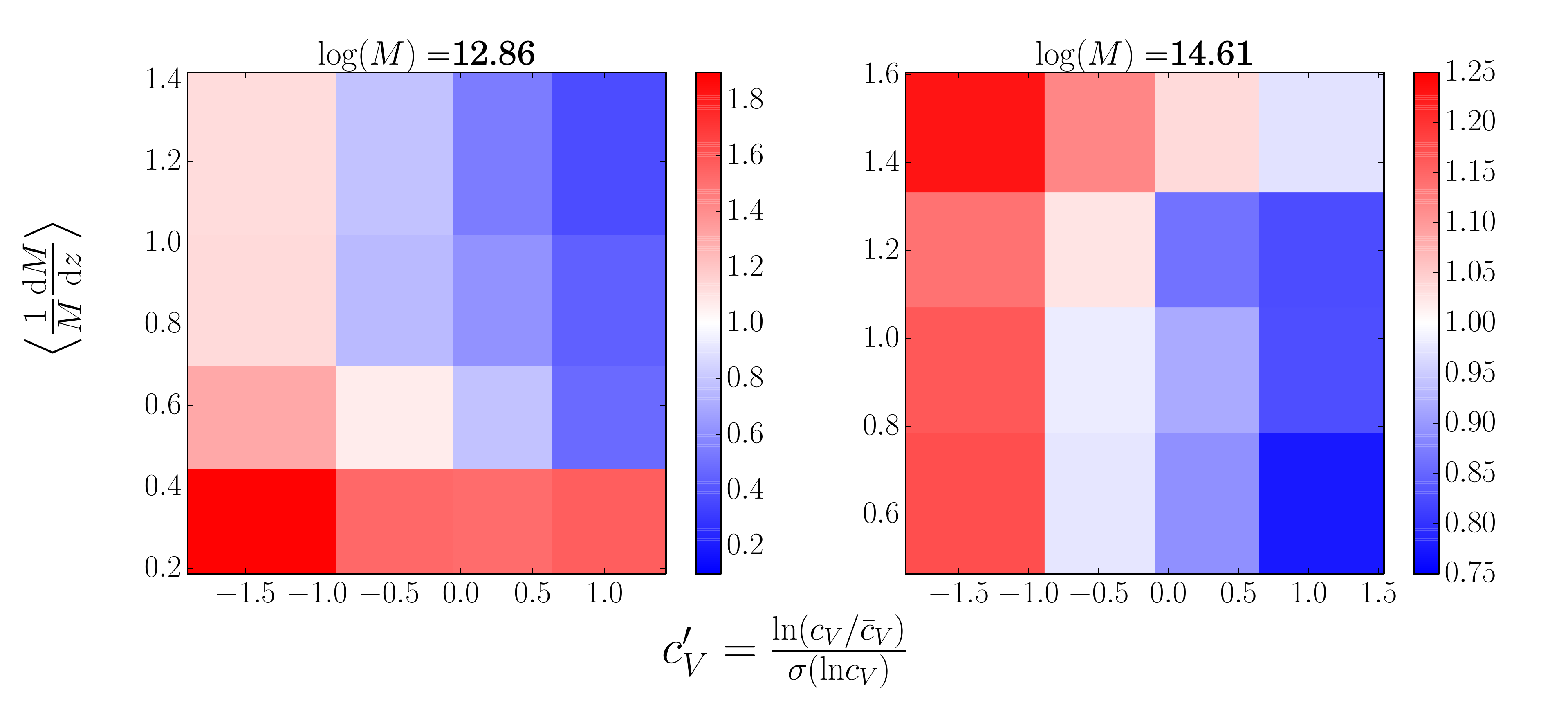}
\caption{$b_1^{\rm rel}$ as a function of logarithmic mass accretion rate and concentration for the highest and lowest mass bins. A white cell corresponds to $b_1^{\rm rel}=1$.}
\label{fig:bofMdotandc}
\end{figure}

\begin{figure}
\centering
\includegraphics[scale=0.37]{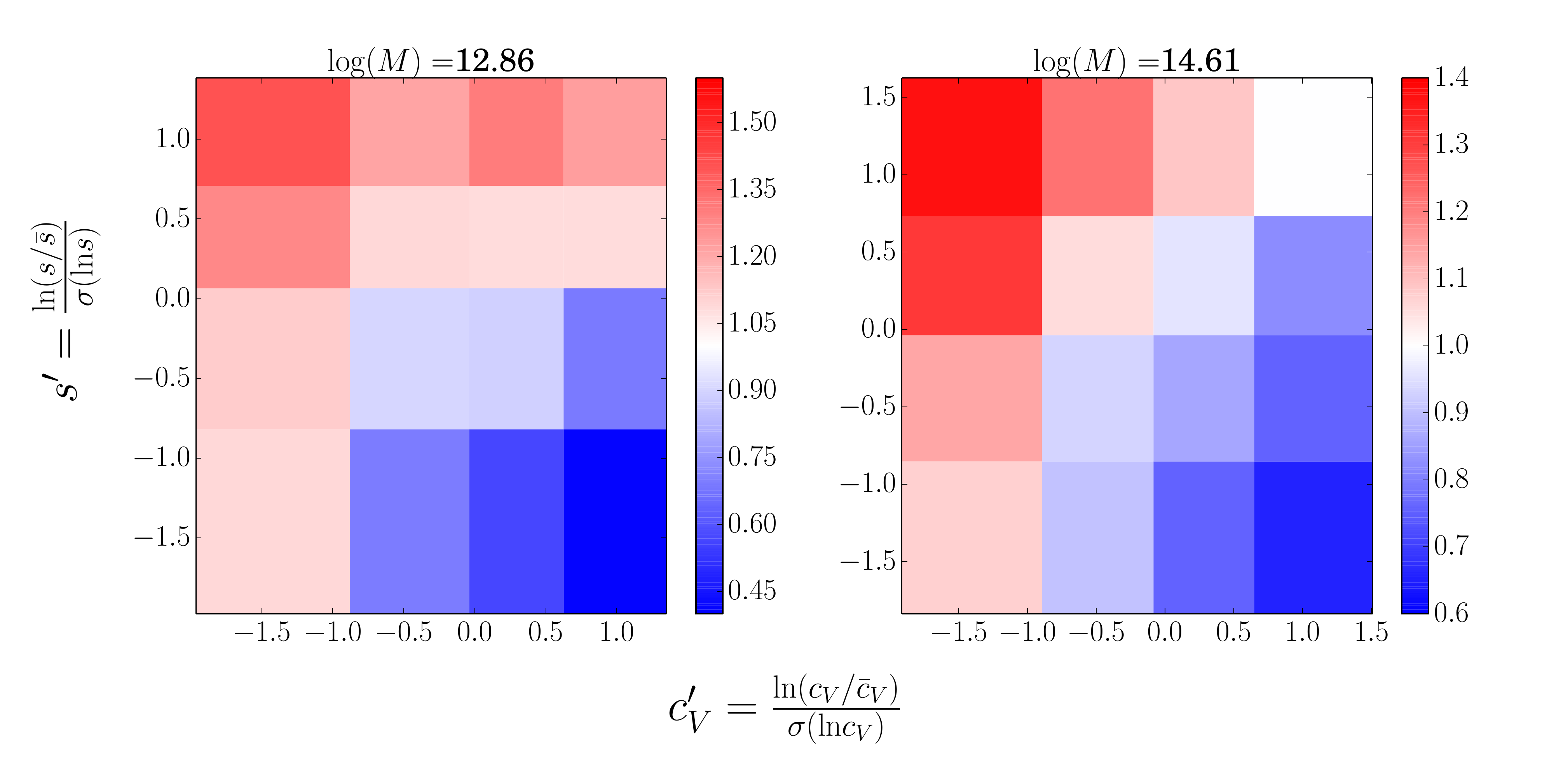}
\caption{$b_1^{\rm rel}$ as a function of shape and concentration for the highest and lowest mass bins. A white cell corresponds to $b_1^{\rm rel}=1$.}
\label{fig:bofsandc}
\end{figure}

\begin{figure}
\centering
\includegraphics[scale=0.42]{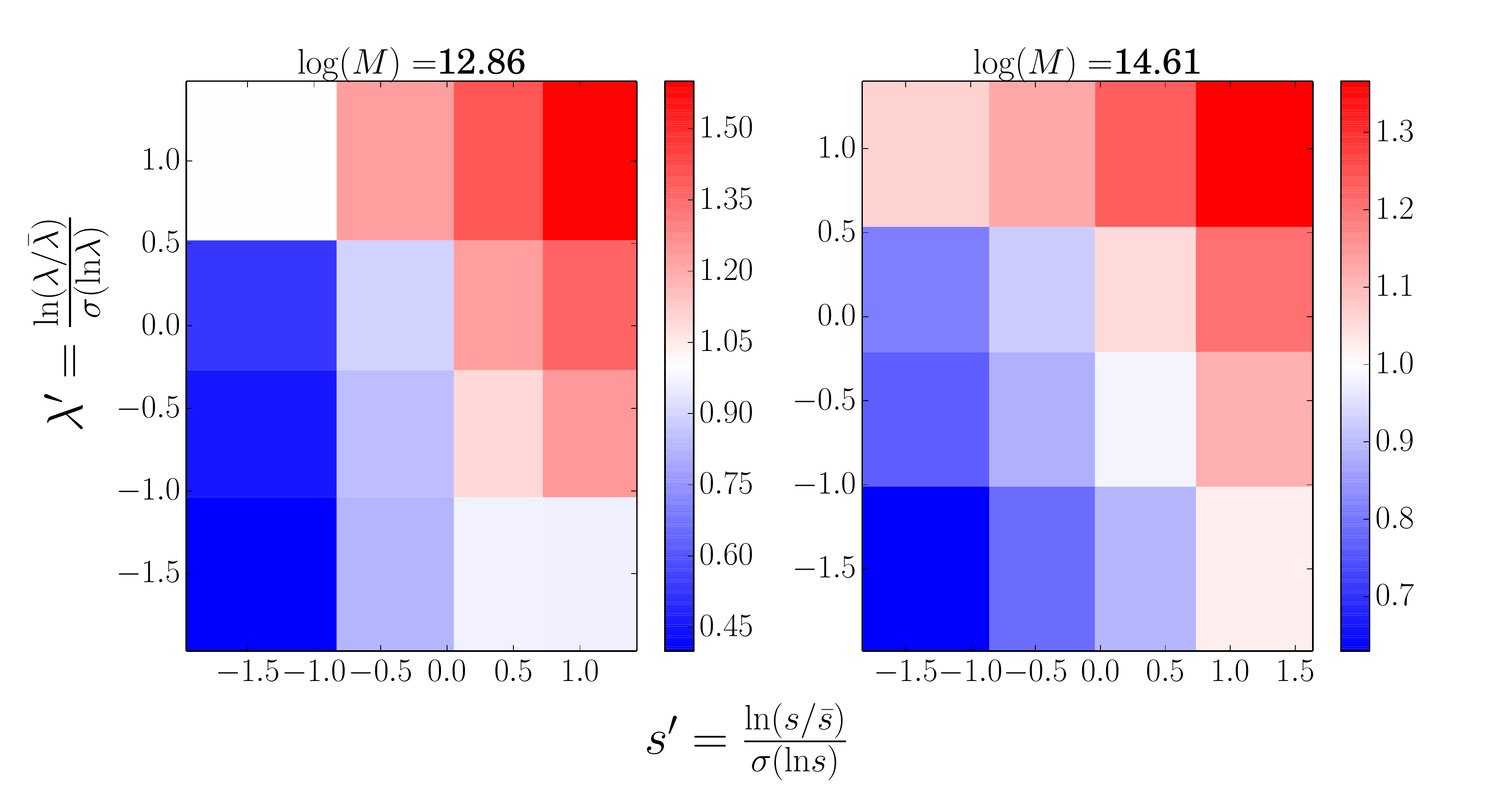}
\caption{$b_1^{\rm rel}$ as a function of spin parameter and shape for the lowest and highest mass bins. A white cell corresponds to $b_1^{\rm rel}=1$.}
\label{fig:bofLands}
\end{figure}

\refFig{bofLandc} presents the dependence of the bias on concentration and spin. As expected from previous results, the bias is maximum at low concentration and high spin parameter, and minimum at high concentration and low angular momentum in both mass bins. Furthermore, the increase in $\bir{1}$ with $\lambda$ is still observed at all mass and concentration. The behaviour of the bias with concentration at all mass and spin is also in agreement with what is shown on \reffig{bofc}. Evidence of a reversal of the trend of $\bir{1}(c)$ at low mass compared to high mass can be seen, especially for the two intermediate spin parameter values. 

Moving to \reffig{bofMdotandc}, one can see that the bias is maximum at low concentration and low mass accretion and minimum for high concentration and high mass accretion, with a mild dependence on $M^{-1}{\rm d}M/{\rm d}z$ at all concentration and mass. Once again this is expected from the results of the previous section (\reffig{bofc} and \reffig{bofMdot}). The dependence of the linear bias on concentration seems slightly enhanced with respect to \reffig{bofc} and \reffig{bofLandc}. The effect is most visible in the lowest mass bin where a clear decrease of $\bir{1}$ with $c_V$ can be observed, instead of the expected plateauing.

\refFig{bofsandc} shows results for $\bir{1}(s,c)$. More spherical and less concentrated halos are more clustered and the opposite behaviour is visible for ellipsoidal, less concentrated halos. In addition to these already known relations of halo bias with $c_V$ and $s$ (\reffig{bofc} and \reffig{bofs}), an interesting effect shown on our plot is that, in the lowest mass bin, the dependence of halo bias on concentration is stronger at low $s$  (more elliptical halos) than high $s$. Equivalently, the opposite is true for $\bir{1}(s)$ with the dependence being stronger at high $c_V$ at all masses. We have observed this behaviour in all mass bins except the highest one.

Finally, \reffig{bofLands} presents results for the relative bias as a function of halo shape and spin parameter. As is again expected from previous results, more spherical halos with higher angular momentum are more clustered than low-spin, more elliptical ones. The dependence of the bias on $s$ is enhanced at fixed spin parameter with respect to \reffig{bofs}. This enhancement is stronger at low mass and low spin parameter. 

To summarize this section, the joint dependence of bias on two properties at fixed mass is in broad agreement with what was found for the dependence on one property at a time in the previous section (\reffigs{bofc}{bofMdot}). Specifying an extra halo property does not change the general trend of $\bir{1}$ with another quantity but sometimes enhances the effect. When it comes to halo shape, this enhancement is more important for elliptical halos than for spherical ones. This would mean that assembly bias with respect to one halo property is at best only mildly correlated with assembly bias with respect to to another halo property. In addition, as already established in the previous section, knowing the assembly bias as a function of one of the two properties in these figures and the mean relation between the two properties would not be enough to fully reconstruct these two dimensional plots. We do not see any clear evidence against the separability of $\bir{1}(p_1,p_2)$ into two independent functions $g(p_1)$ and $h(p_2)$. We however have not ruled this out rigorously as this would require a careful statistical analysis.

%%%%%%%%%%%%%%%%%%%%%%%%%%%%%%%%%%%%%%%%%%%%%%%%%%%%%%%%%%%%%%%%%%%%%%%%%%%%%
%%%%%%%%%%%%%%%%%%%%%%%%%%%%%%%%%%%%%%%%%%%%%%%%%%%%%%%%%%%%%%%%%%%%%%%%%%%%%
\section{Conclusions}
\label{sec:concl} 

We have presented new measurements of the assembly bias of dark matter halos using separate universe simulations. Before drawing our conclusions and outlook, we recap the main points of our results:
\begin{itemize}
\item The separate universe approach allows us to measure the assembly bias precisely in the \textit{large-scale limit}, in contrast with previous studies, almost all of which used the halo correlation function on scales of $20 h^{-1} {\rm Mpc}$ or less.  Strictly speaking, these are the renormalized biases that enter the perturbation theory prediction for large-scale halo $n$-point functions.
\item We have obtained the first measurements of large-scale assembly bias in $b^E_2$ (see also \cite{Paranjape:2016}, who recently measured $b^E_2(c)$ using the same technique). The trends in $b^E_2$ are the same as those in $b^E_1$ for all halo properties and at all masses. 
\item We present the first measurements of assembly bias with respect to the late-time mass accretion rate. 
\item We present the first measurements of assembly bias with respect to two halo properties simultaneously.
\end{itemize}

Concerning our results for the linear bias, we found good agreement overall with previous works, where available, both qualitatively and quantitatively. The good quantitative agreement was quite surprising since the scales considered are quite different. However, this could be due to the fact that the trends in $b^E_2$ go in the same direction as those in $b^E_1$ for all halo properties. In addition, our error bars are much smaller than previous ones (e.g. \reffig{bofcwechs}). This shows again the power of the separate universe technique already highlighted in \cite{Lazeyras:2015}.

Another important result was the obvious impossibility to reconstruct the relative linear bias as a function of property $p_1$ using the assembly bias with respect to another property $p_2$ and the relation $p_1(p_2)$ (see \reffigs{bofc}{bofMdot}). This was an already known fact. However, we showed that some combinations work better than others. For example, $\lambda$ and $c_V$ can be used to roughly reconstruct the bias with respect to each other (at least qualitatively). The shape $s$ on the other hand works very poorly. We interpret this as the fact that highly biased halos (at fixed mass) do not all belong to the same population. The populations of high angular momentum and low concentration halos seem to have substantial overlap (which could be due to a more extended mass distribution of a halo being associated with higher angular momentum), but do not match the population of roughly spherical halos. To make the study more complete, it would be interesting to conduct a principal component analysis (PCA) such as the one presented in \cite{Skibba:2011}. This however goes beyond the scope of this work. 

The plots of assembly bias with respect to two halo properties (\refsec{2bin}) are in agreement with what one could expect from the results of \refsec{abiasres}, in the sense that specifying an extra halo property does not change the general trend of $\bir{1}$ with another quantity. This confirms that assembly bias with respect to one halo property is only mildly correlated with assembly bias as a function of another one, as already shown by the reconstructed curves. As we already stated, we do not see any evidence against the separability of $\bir{1}(p_1,p_2)$ into two independent functions $g(p_1)$ and $h(p_2)$ although we did not do not prove this rigorously.

We also investigated how assembly bias can arise in the ESP model, studying the dependence of the bias on the stochastic variable $\beta$ and the mass accretion rate. A higher $\beta$ implies a higher threshold for collapse and hence a higher bias. We interpret this as the effect of the initial shear making halos more elliptical. These halos then necessitate a higher internal density to collapse. However, we showed that linking this interpretation to late time halo shape (or any other anisotropy parameters) is nontrivial, as the behaviour of the linear bias as a function of $\beta$ is inverse to the one with respect to final halo shape $s$.  A more detailed comparison is possible in case of the mass accretion rate. We found qualitative agreement between the ESP prediction and our measurements, especially for lower halo masses.  The quantitative agreement is very poor at all masses, which is expected given that negative mass accretion rates are impossible in the excursion set picture, while real halos clearly do show mass loss (see \refapp{massev}).  

Finally, significant interest has developed lately for looking at halo properties as a function of their final environment (see e.g. \cite{Borzyszkowski:2016,Lee:2016} and references therein) in order to shed new lights on assembly bias. While this is certainly of crucial importance as the environment of a halo drives its evolution and, hence, determines its internal final properties, it is not clear that late-time environment variables (such as the shear or the position in the cosmic web) are enough to fully explain assembly bias. As shown in e.g. \cite{Borzyszkowski:2016}, quantities such as the initial shear can play an important role in halo formation. It would thus be interesting to push investigations further in this direction in order to better link properties of protohalos, as well as their environment, to late time evolution parameters. 
One open question that still remains is to establish if a finite set of halo properties are sufficient to describe the assembly bias of dark matter halos, and, if so, how many and what these properties are.

\acknowledgments{We thank S. White and S. More for useful comments regarding our measurements. We further thank R. Sheth for insightful comments and discussions. MM thanks the University of Pennsylvania for support during the early stage of this work.
FS acknowledges support from the Marie Curie Career Integration Grant  (FP7-PEOPLE-2013-CIG) ``FundPhysicsAndLSS,'' and Starting Grant (ERC-2015-STG 678652) ``GrInflaGal'' from the European Research Council.
}

\FloatBarrier

%%%%%%%%%%%%%%%%%%%%%%%%%%%%%%%%%%%%%%%%%%%%%%%%%%%%%%%%%%%%%%%%%%%%%%%%%%%%%
%%%%%%%%%%%%%%%%%%%%%%%%%%%%%%%%%%%%%%%%%%%%%%%%%%%%%%%%%%%%%%%%%%%%%%%%%%%%%
\appendix

\section{Comparison of $\bie{1}$ with W06}
\label{app:wechsler} 

\begin{figure}
\centering
\includegraphics[scale=0.35]{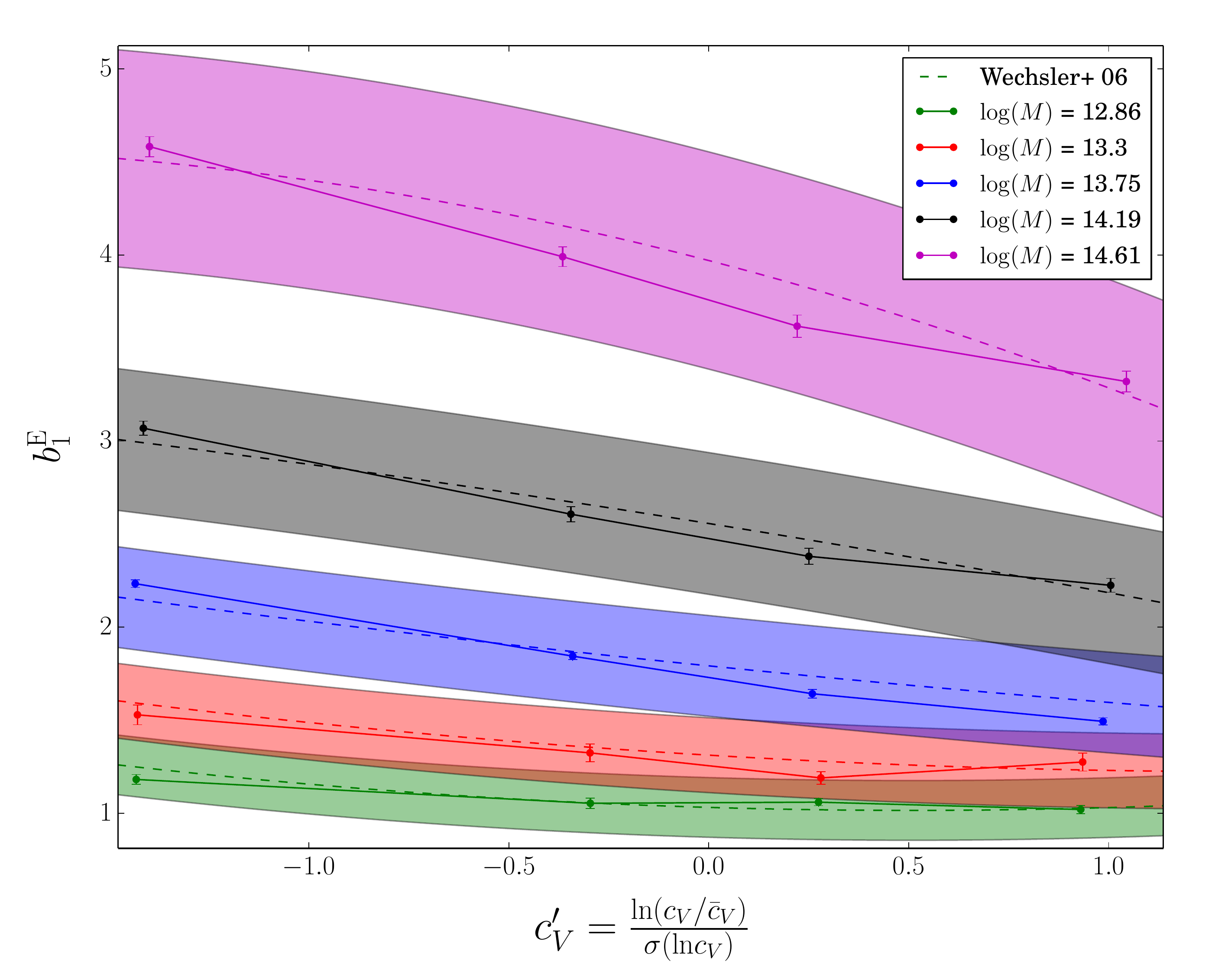}
\caption{Eulerian linear bias as a function of concentration for several mass bins (color coded as on \reffigs{bofc}{bofMdot}). The dotted lines and shaded areas represent the best fit from W06 and an estimation of the corresponding $1 \sigma$ error from their figure 4.}
\label{fig:bofcwechs}
\end{figure}

This appendix presents a quantitative comparison of our results with the best fit of W06. \refFig{bofcwechs} shows the Eulerian linear bias as a function of concentration for several mass bins. The points linked by solid lines are the measurements from this work while the dashed curves are the best fit of W06. The shaded regions represent an estimate of the $1\sigma$ errors on their figure 4. As can be seen, our measurements are well within their error bars for all masses but our $1\sigma$ errors are much smaller than theirs, showing once again the power of the separate universe simulations technique to infer precise measurements of the large-scale dark matter halo bias as already emphasized in \cite{Lazeyras:2015}. See the main text of \refsec{abiasres} for a more detailed discussion of this figure.

\section{Relations between halo properties}
\label{app:haloprop}

In this appendix, we present numerous plots of mean relations between two halo properties. These relations are obtained from scatter plots by first binning in a halo property $p_1$ (shown on the $x$ axes) using the same bins as for the main results (i.e quartiles of the total distribution in the fiducial cosmology) and computing the mean of another property $p_2$ (shown on the $y$ axes) in these bins. We use these plots to infer assembly bias with respect to $p_1$ using the assembly bias with respect to $p_2$.  As stated in the main text (\refsec{abiasres}), the results obtained in this way in general agree poorly with the direct measurements of assembly bias with respect to $p_1$. We tried to obtain theses relations by a direct fit of the full point cloud but this did not make the agreement better, even qualitatively. This known failure shows that none of the four halo properties considered are able to explain the entire assembly bias phenomenon, for any of the mass bins considered here.

\begin{figure}
\centering
\includegraphics[scale=0.47]{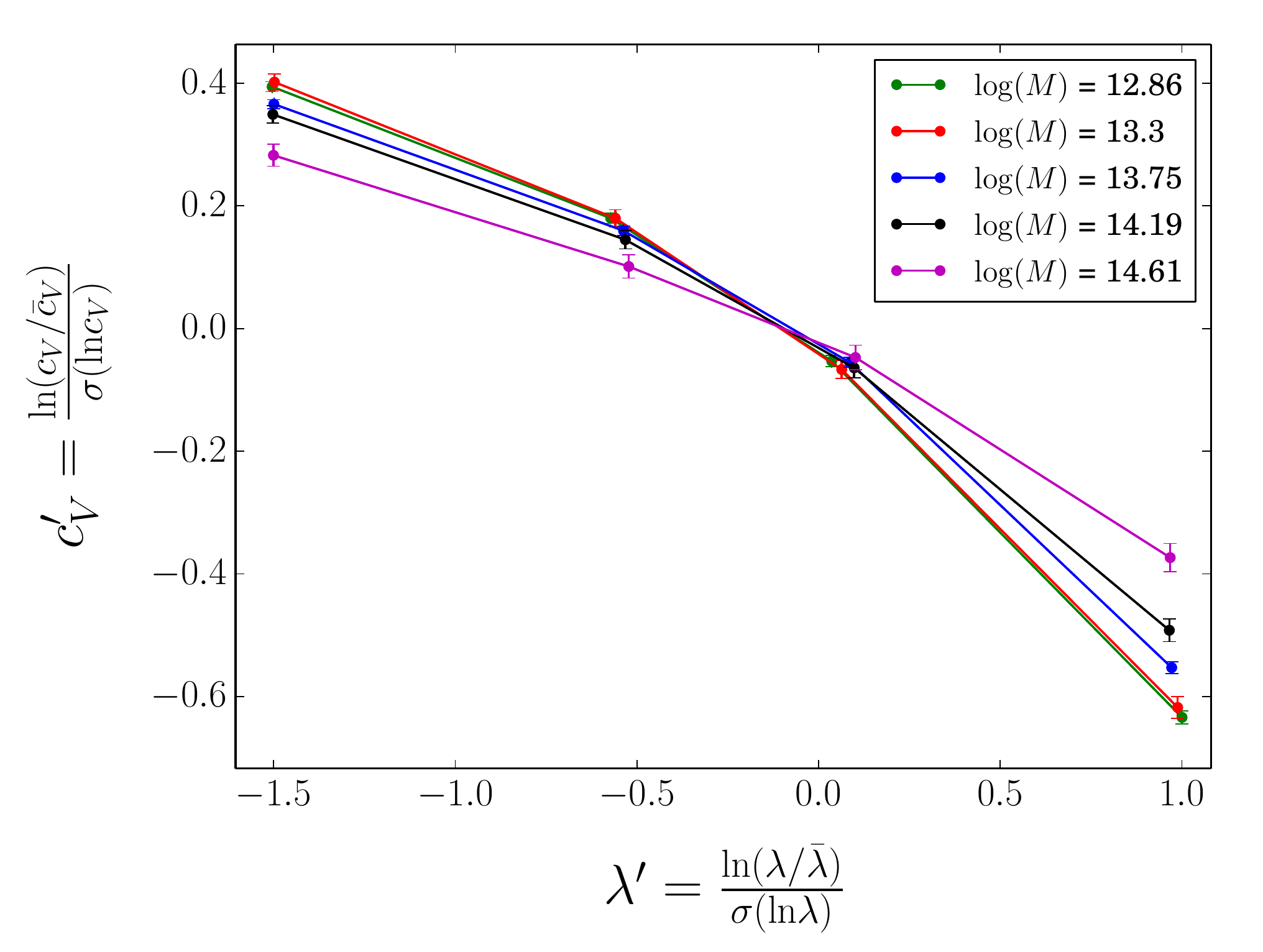}
\caption{Mean relation between halo concentration and spin for each mass bin. The color coding follows the one in \reffigs{bofc}{bofMdot}.}
\label{fig:cofL}
\end{figure}

\begin{figure}
\centering
\includegraphics[scale=0.47]{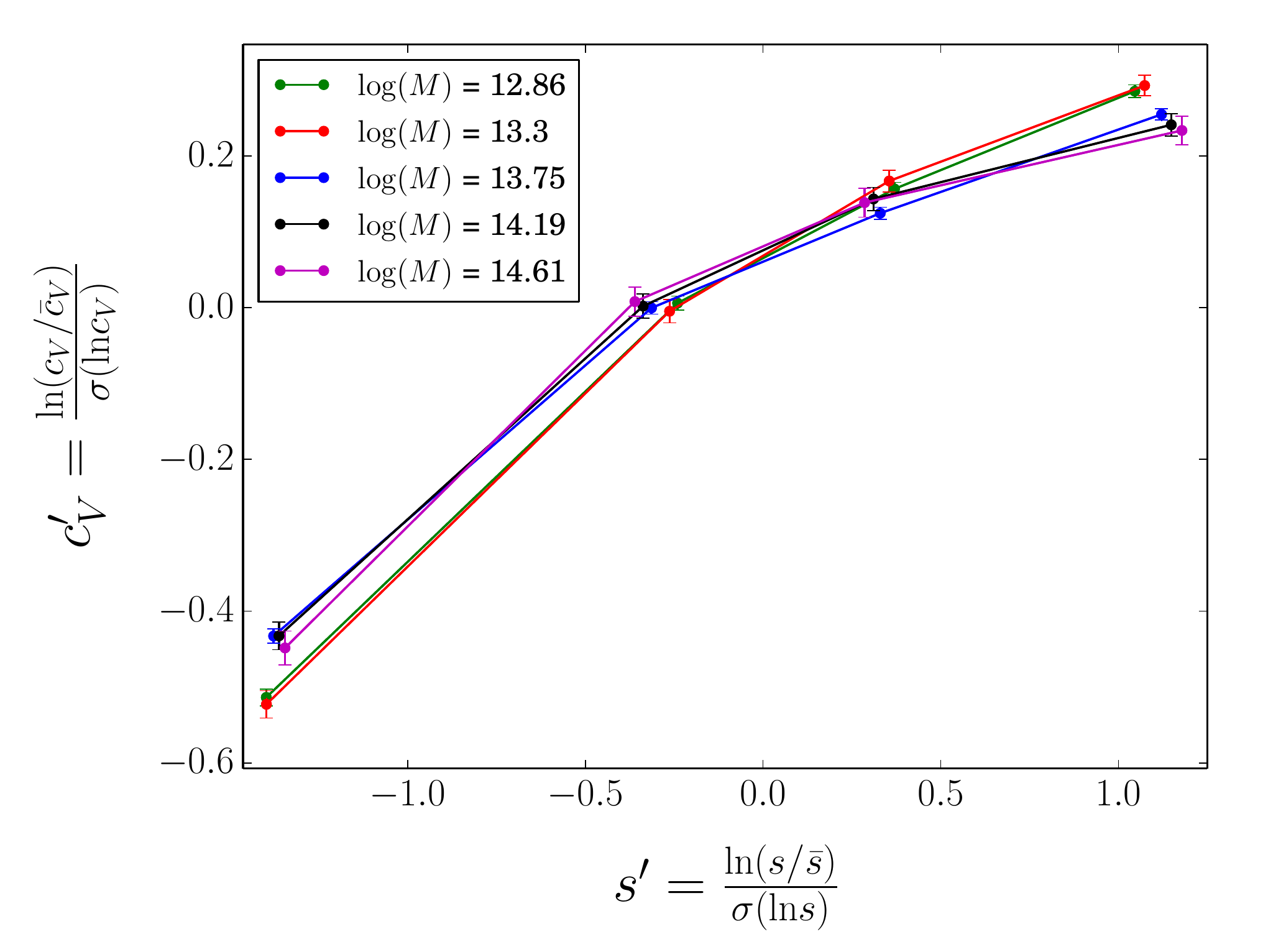}
\caption{Same as \reffig{cofL} but for the concentration as a function of shape.}
\label{fig:cofs}
\end{figure}

\begin{figure}
\centering
\includegraphics[scale=0.47]{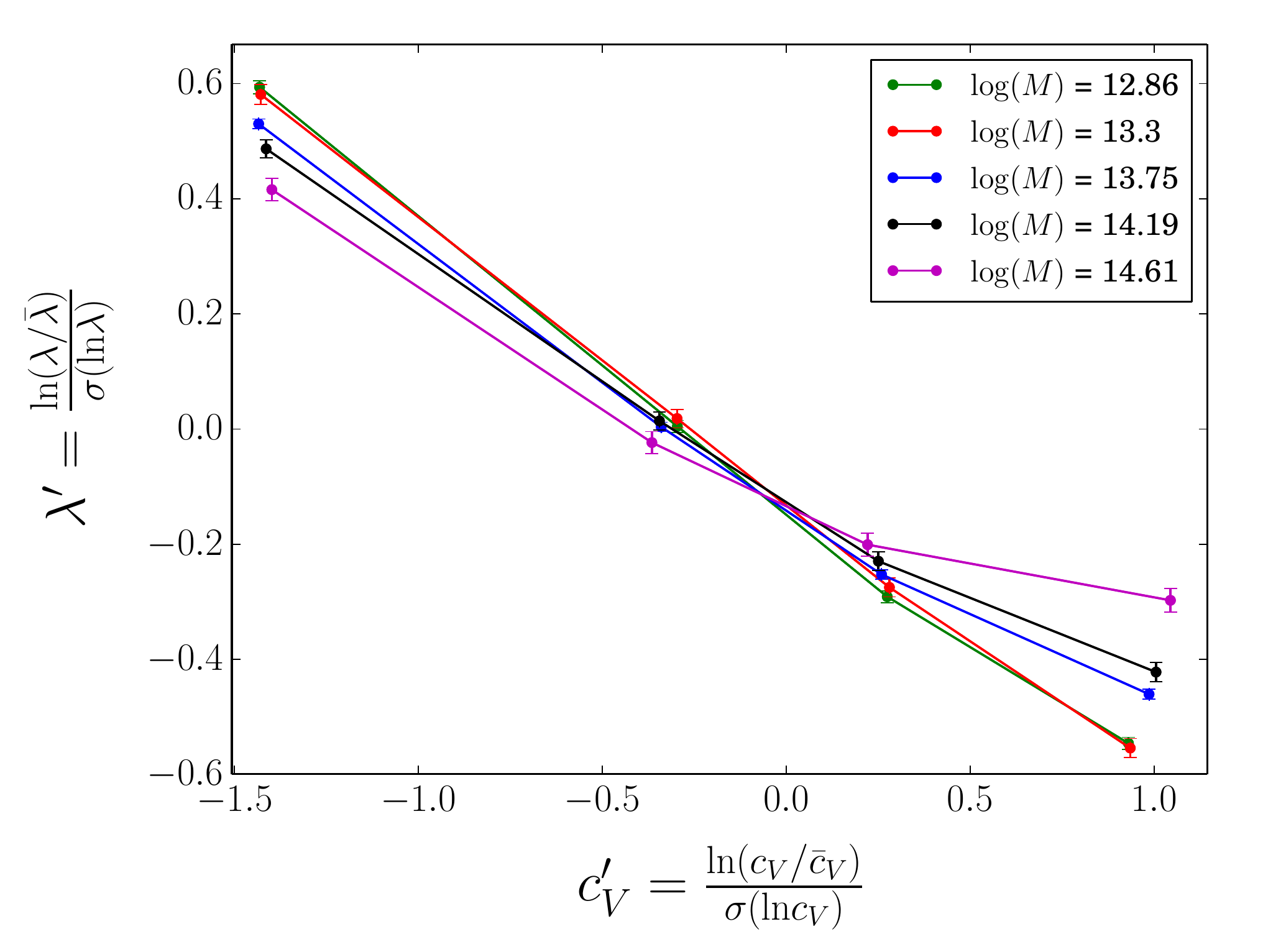}
\caption{Same as \reffig{cofL} but for the spin parameter as a function of concentration.}
\label{fig:Lofc}
\end{figure}

\begin{figure}
\centering
\includegraphics[scale=0.47]{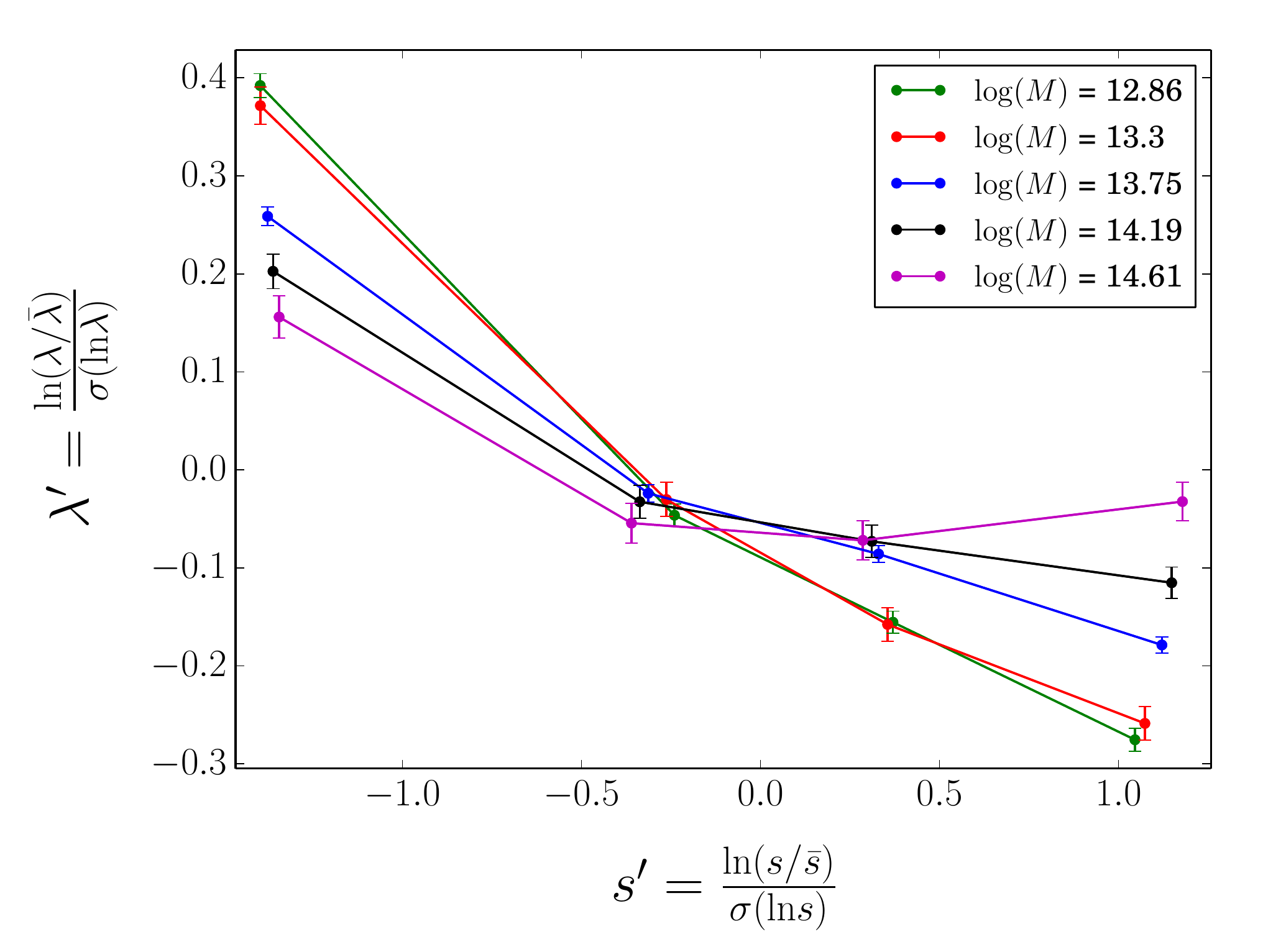}
\caption{Same as \reffig{cofL} but for the spin parameter as a function of shape.}
\label{fig:Lofs}
\end{figure}

\begin{figure}
\centering
\includegraphics[scale=0.47]{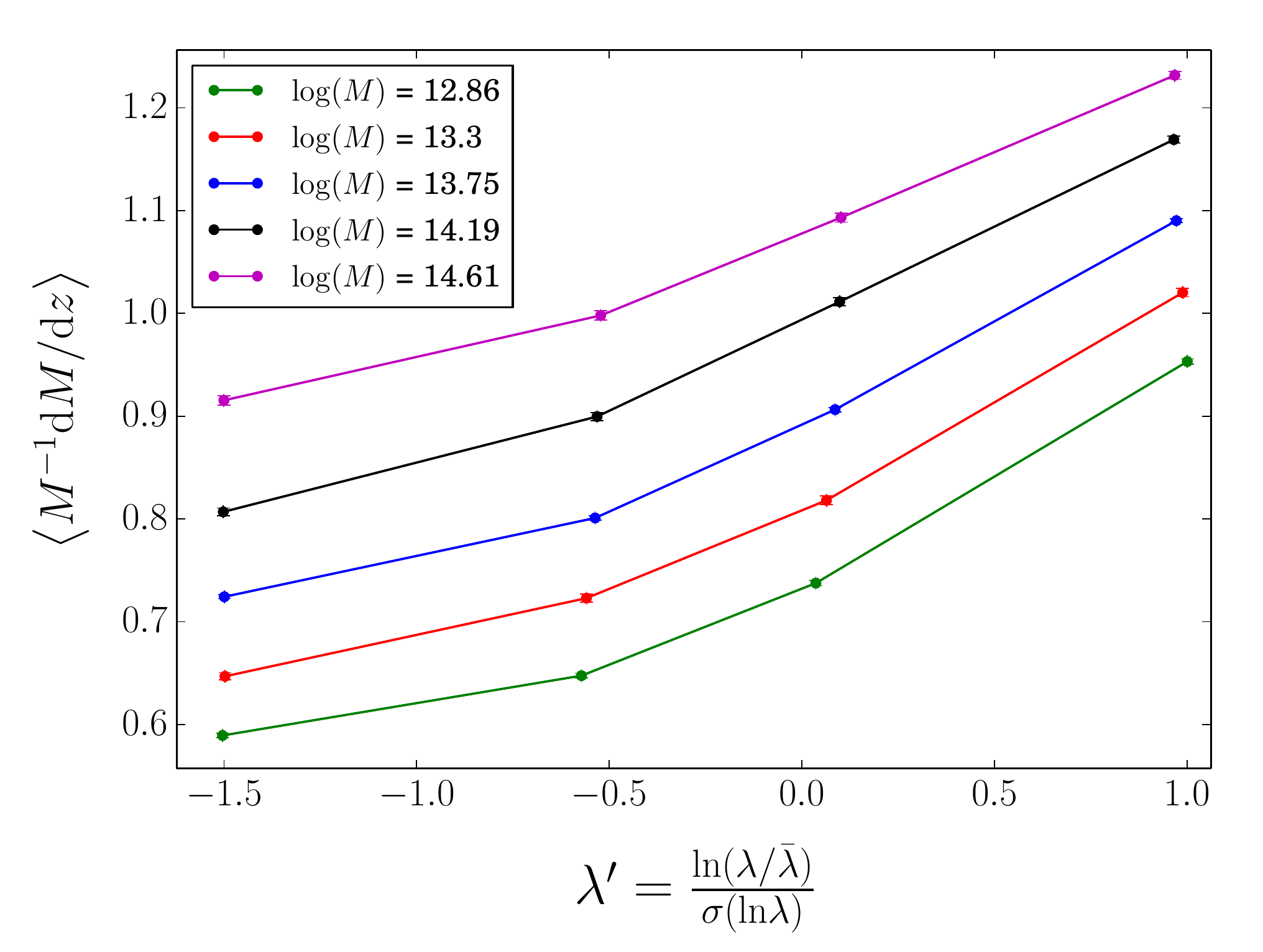}
\caption{Same as \reffig{cofL} but for the mass accretion rate as a function of spin parameter.}
\label{fig:MdotofL}
\end{figure}

\begin{figure}
\centering
\includegraphics[scale=0.47]{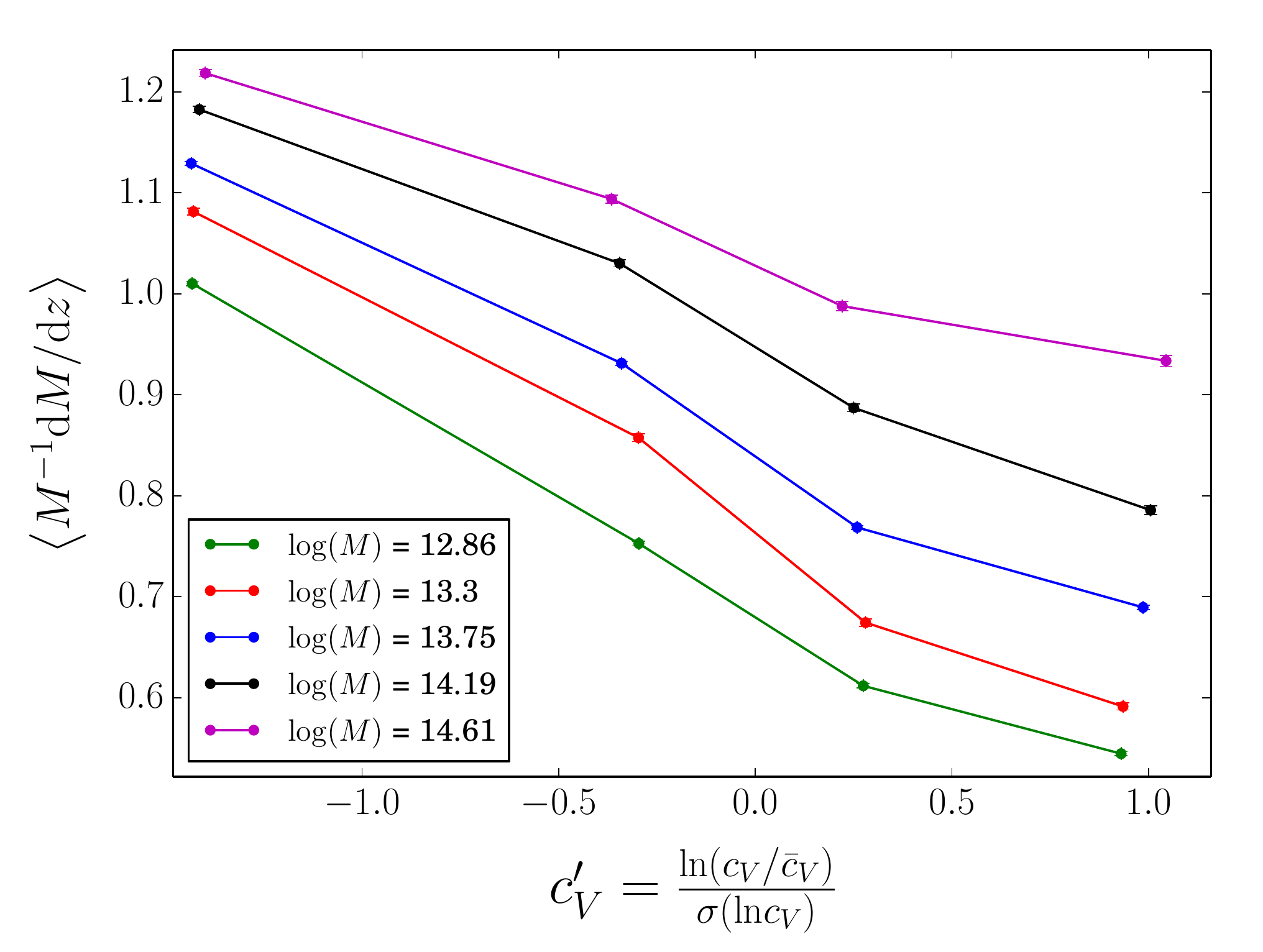}
\caption{Same as \reffig{cofL} but for the mass accretion rate as a function of concentration.}
\label{fig:Mdotofc}
\end{figure}

\begin{figure}
\centering
\includegraphics[scale=0.47]{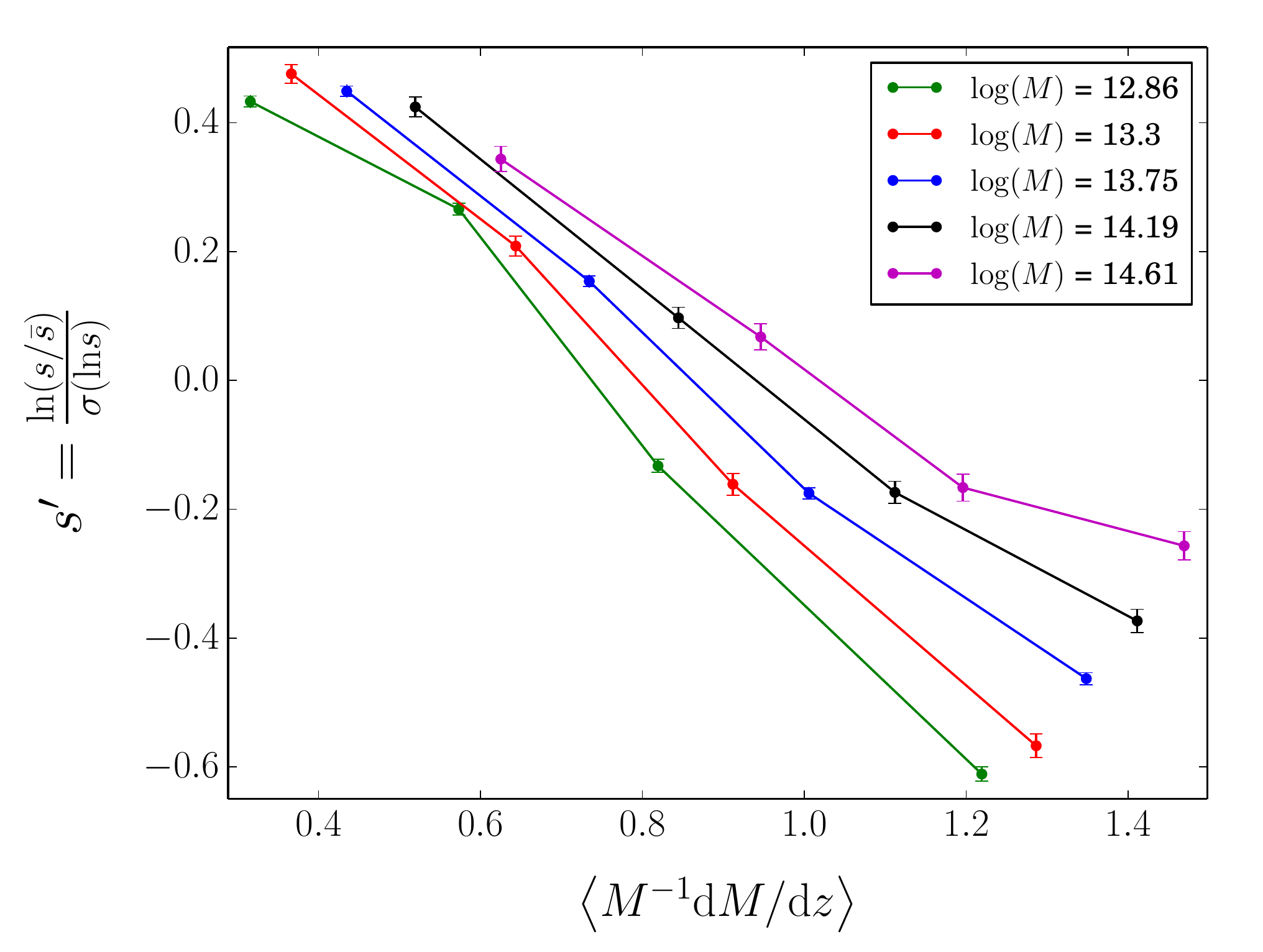}
\caption{Same as \reffig{cofL} but for the shape as a function of mass accretion rate.}
\label{fig:sofMdot}
\end{figure}

\begin{figure}
\centering
\includegraphics[scale=0.47]{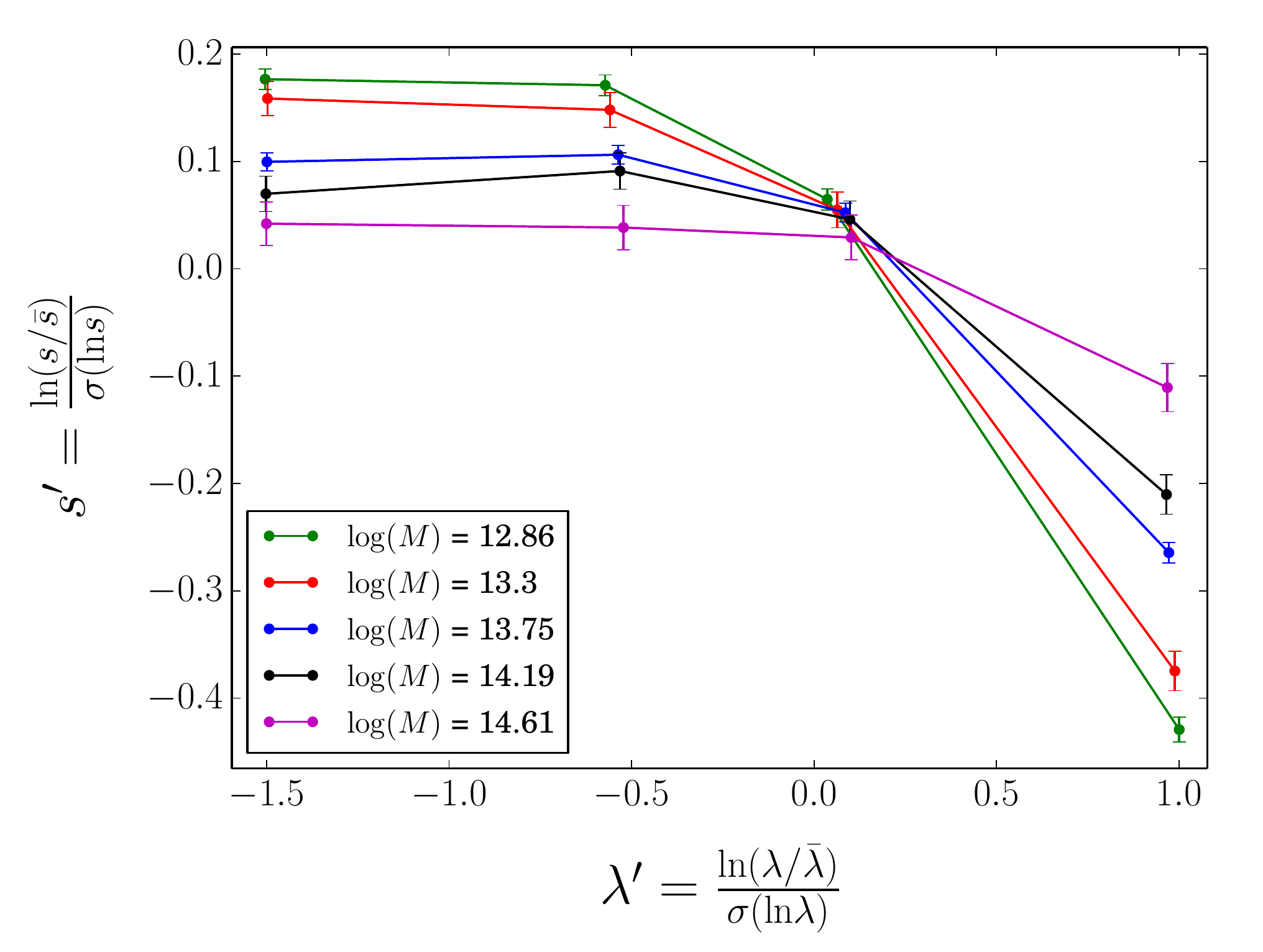}
\caption{Same as \reffig{cofL} but for the shape as a function of spin parameter.}
\label{fig:sofL}
\end{figure}

\FloatBarrier

\section{Mass evolution of the mass accretion rate}
\label{app:massev}

\begin{figure}[h]
\centering
\includegraphics[trim=0cm 0.3cm 0cm 0cm,clip,width=0.43\textwidth]{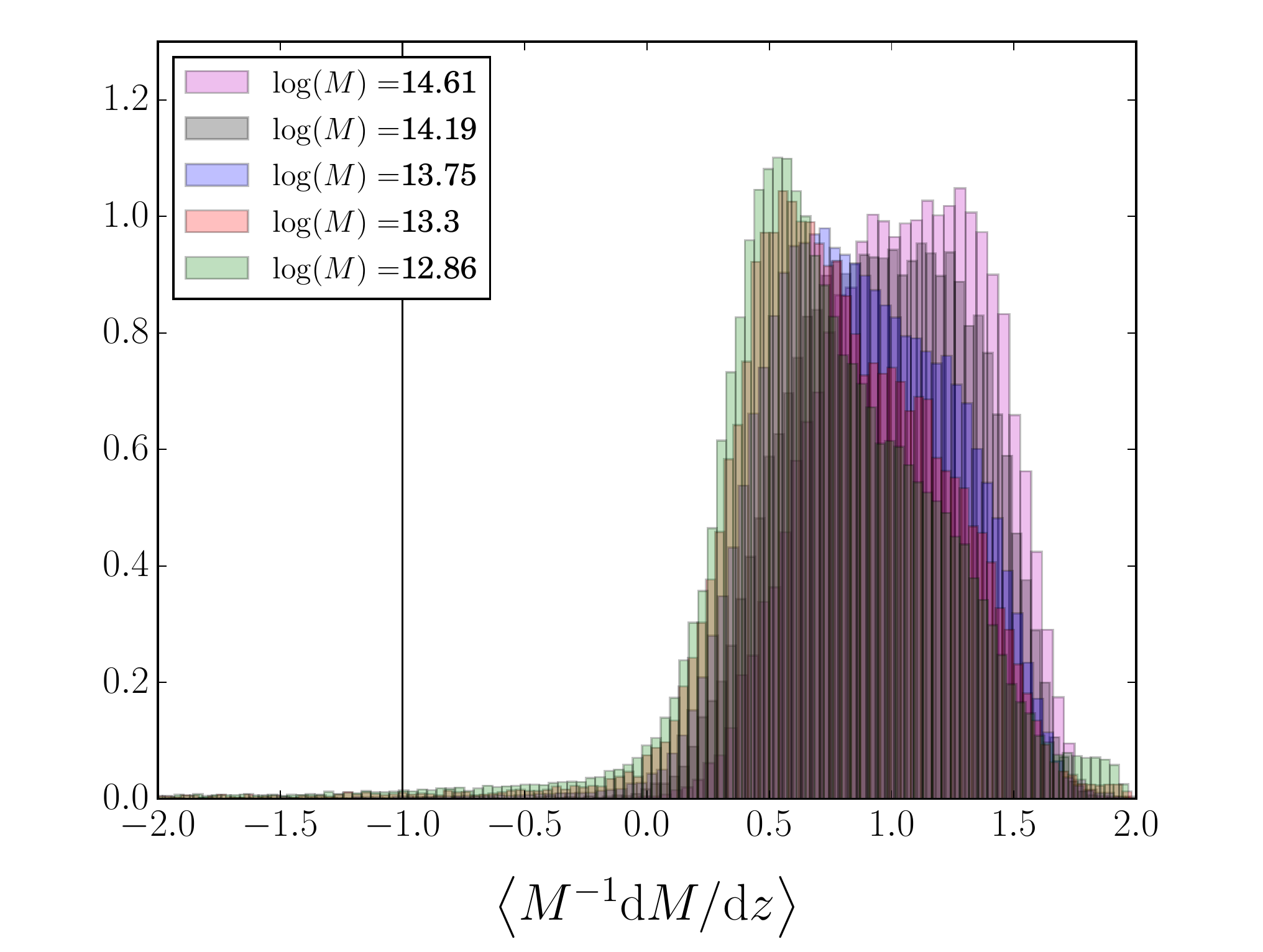} 
\includegraphics[trim=0.0cm 0.0cm 0.2cm 0.7cm,clip,width=0.5\textwidth]{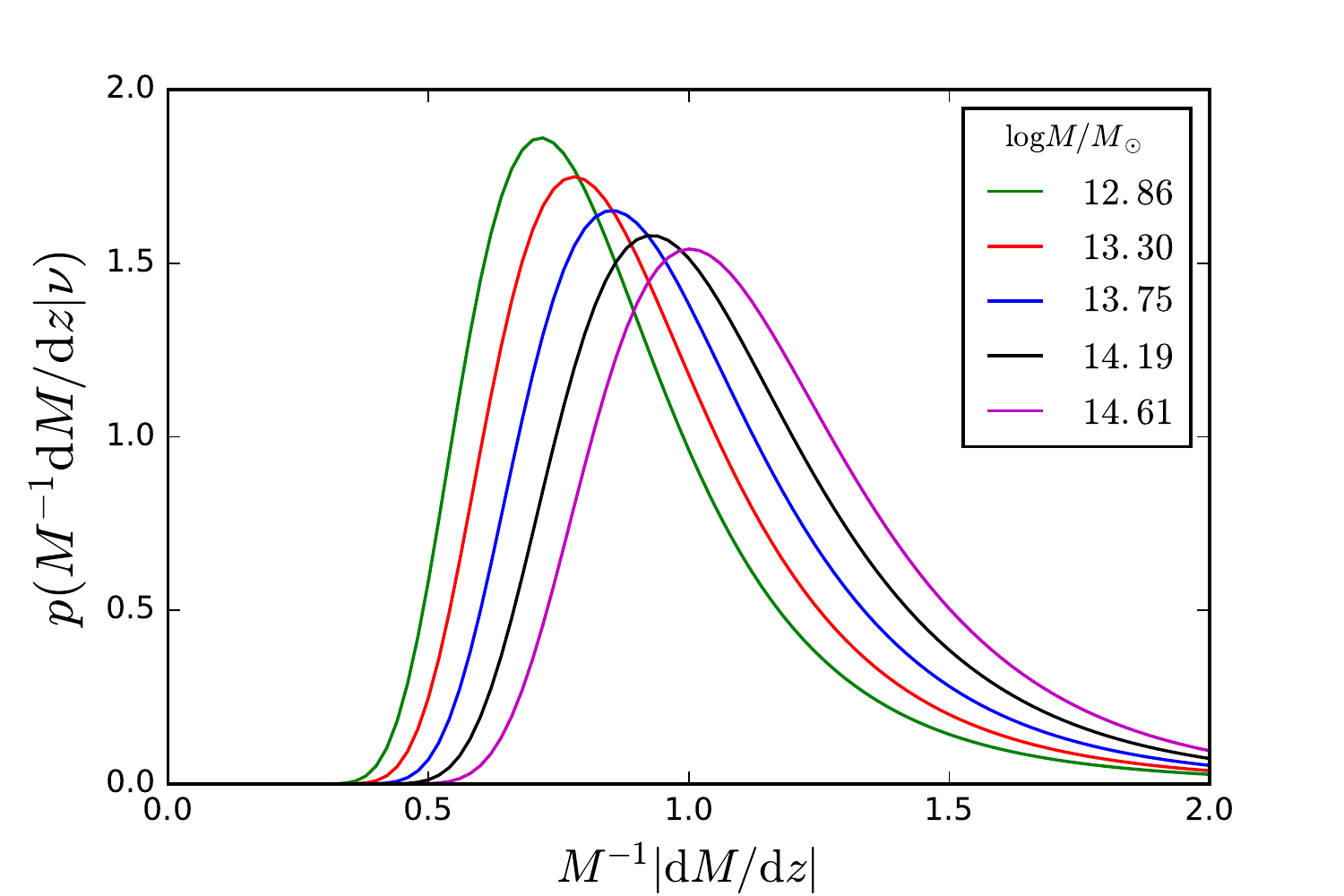} 
\caption{\textbf{Left panel: }Measured mass evolution of the normalized distribution of the mass accretion rate of halos. The vertical black line represents our cut-off for maximal negative mass accretion. \textbf{Right panel: }Evolution of the normalized probability density function of the same quantity as predicted by the ESP. One immediately sees that negative accretion rates are forbidden.}
\label{fig:Mdothist}
\end{figure}

\refFig{Mdothist} presents the mass evolution of the mass accretion rate distribution from simulations (left panel) and as predicted by the ESP (right panel). The color coding follows that of \reffigs{bofc}{bofMdot}. As one can see, the mass accretion rate monotonically increases with increasing mass. In addition, we use this plot to justify our choice of putting a lower limit of -1 for $M^{-1} {\rm d}M/{\rm d}z$ (vertical line). Indeed, this cut allows us to discard very negative mass accretion rates without neglecting a large fraction of the distribution. 

The reason we do not want to consider these halos to infer assembly bias as a function of mass accretion rate is that they most likely correspond to strongly tidally stripped halos that recently passed through a much larger halo without having merged yet. These satellite halos hence lie in the vicinity of much larger structures with which they will eventually merge and are wrongly identified as main structures by the halo finder. As their spatial distribution by definition follows that of high mass halos, they are highly clustered, which could artificially enhance the effect of assembly bias. 

Notice however that allowing for, albeit moderately, negative mass accretion rates represents a qualitative difference from the ESP which only allows for monotonically increasing halo mass.  This could clearly lead to disagreements between the theoretical prediction and the measurements. 

%%%%%%%%%%%%%%%%%%%%%%%%%%%%%%%%%%%%%%%%%%%%%%%%%%%%%%%%%%%%%%%%%%%%%%%%%%%%
%%%%%%%%%%%%%%%%%%%%%%%%%%%%%%%%%%%%%%%%%%%%%%%%%%%%%%%%%%%%%%%%%%%%%%%%%%%%

\bibliography{references}

\providecommand{\href}[2]{#2}\begingroup\raggedright\begin{thebibliography}{10}

\bibitem{Desjacques:2016}
V.~Desjacques, D.~Jeong, and F.~Schmidt, {\it {Large-Scale Galaxy Bias}},
  \href{http://arxiv.org/abs/1611.09787}{{\tt arXiv:1611.09787}}.

\bibitem{mcdonald}
P.~{McDonald}, {\it {Clustering of dark matter tracers: Renormalizing the bias
  parameters}},  {\em \prd} {\bf 74} (Nov., 2006) 103512,
  [\href{http://arxiv.org/abs/astro-ph/0}{{\tt astro-ph/0}}].

\bibitem{assassi/etal}
V.~{Assassi}, D.~{Baumann}, D.~{Green}, and M.~{Zaldarriaga}, {\it
  {Renormalized halo bias}},  {\em \jcap} {\bf 8} (Aug., 2014) 56,
  [\href{http://arxiv.org/abs/1402.5916}{{\tt arXiv:1402.5916}}].

\bibitem{PBSpaper}
F.~{Schmidt}, D.~{Jeong}, and V.~{Desjacques}, {\it {Peak-background split,
  renormalization, and galaxy clustering}},  {\em \prd} {\bf 88} (July, 2013)
  023515, [\href{http://arxiv.org/abs/1212.0868}{{\tt arXiv:1212.0868}}].

\bibitem{Bond:1990}
J.~R. Bond, S.~Cole, G.~Efstathiou, and N.~Kaiser, {\it {Excursion set mass
  functions for hierarchical Gaussian fluctuations}},  {\em Astrophys. J.} {\bf
  379} (1991) 440.

\bibitem{Lacey:1993}
C.~G. Lacey and S.~Cole, {\it {Merger rates in hierarchical models of galaxy
  formation}},  {\em Mon. Not. Roy. Astron. Soc.} {\bf 262} (1993) 627--649.

\bibitem{mo/white:1996}
H.~J. {Mo} and S.~D.~M. {White}, {\it {An analytic model for the spatial
  clustering of dark matter haloes}},  {\em \mnras} {\bf 282} (Sept., 1996)
  347--361, [\href{http://arxiv.org/abs/astro-ph/9}{{\tt astro-ph/9}}].

\bibitem{Kauffmann:1995}
G.~Kauffmann, A.~Nusser, and M.~Steinmetz, {\it {Galaxy formation and large
  scale bias}},  {\em Mon. Not. Roy. Astron. Soc.} {\bf 286} (1997) 795--811,
  [\href{http://arxiv.org/abs/astro-ph/9512009}{{\tt astro-ph/9512009}}].

\bibitem{Berlind:2002}
A.~A. Berlind, D.~H. Weinberg, A.~J. Benson, C.~M. Baugh, S.~Cole, R.~Dave,
  C.~S. Frenk, A.~Jenkins, N.~Katz, and C.~G. Lacey, {\it {The Halo occupation
  distribution and the physics of galaxy formation}},  {\em Astrophys. J.} {\bf
  593} (2003) 1--25, [\href{http://arxiv.org/abs/astro-ph/0212357}{{\tt
  astro-ph/0212357}}].

\bibitem{Yang:2002}
X.-h. Yang, H.~J. Mo, and F.~C. van~den Bosch, {\it {Constraining galaxy
  formation and cosmology with the conditional luminosity function of
  galaxies}},  {\em Mon. Not. Roy. Astron. Soc.} {\bf 339} (2003) 1057,
  [\href{http://arxiv.org/abs/astro-ph/0207019}{{\tt astro-ph/0207019}}].

\bibitem{Zheng:2004}
Z.~Zheng, A.~A. Berlind, D.~H. Weinberg, A.~J. Benson, C.~M. Baugh, S.~Cole,
  R.~Dave, C.~S. Frenk, N.~Katz, and C.~G. Lacey, {\it {Theoretical models of
  the halo occupation distribution: Separating central and satellite
  galaxies}},  {\em Astrophys. J.} {\bf 633} (2005) 791--809,
  [\href{http://arxiv.org/abs/astro-ph/0408564}{{\tt astro-ph/0408564}}].

\bibitem{Sheth:2004}
R.~K. Sheth and G.~Tormen, {\it {On the environmental dependence of halo
  formation}},  {\em Mon. Not. Roy. Astron. Soc.} {\bf 350} (2004) 1385,
  [\href{http://arxiv.org/abs/astro-ph/0402237}{{\tt astro-ph/0402237}}].

\bibitem{gao:2005}
L.~Gao, V.~Springel, and S.~D.~M. White, {\it {The Age dependence of halo
  clustering}},  {\em Mon. Not. Roy. Astron. Soc.} {\bf 363} (2005) L66--L70,
  [\href{http://arxiv.org/abs/astro-ph/0506510}{{\tt astro-ph/0506510}}].

\bibitem{Gao:2006}
L.~Gao and S.~D.~M. White, {\it {Assembly bias in the clustering of dark matter
  haloes}},  {\em Mon. Not. Roy. Astron. Soc.} {\bf 377} (2007) L5--L9,
  [\href{http://arxiv.org/abs/astro-ph/0611921}{{\tt astro-ph/0611921}}].

\bibitem{Wechsler:2005}
R.~H. Wechsler, A.~R. Zentner, J.~S. Bullock, and A.~V. Kravtsov, {\it {The
  dependence of halo clustering on halo formation history, concentration, and
  occupation}},  {\em Astrophys. J.} {\bf 652} (2006) 71--84,
  [\href{http://arxiv.org/abs/astro-ph/0512416}{{\tt astro-ph/0512416}}].

\bibitem{Jing:2006}
Y.~P. Jing, Y.~Suto, and H.~J. Mo, {\it {The dependence of dark halo clustering
  on the formation epoch and the concentration parameter}},  {\em Astrophys.
  J.} {\bf 657} (2007) 664--668,
  [\href{http://arxiv.org/abs/astro-ph/0610099}{{\tt astro-ph/0610099}}].

\bibitem{Croton:2006}
D.~J. Croton, L.~Gao, and S.~D.~M. White, {\it {Halo assembly bias and its
  effects on galaxy clustering}},  {\em Mon. Not. Roy. Astron. Soc.} {\bf 374}
  (2007) 1303--1309, [\href{http://arxiv.org/abs/astro-ph/0605636}{{\tt
  astro-ph/0605636}}].

\bibitem{Angulo:2007}
R.~E. Angulo, C.~M. Baugh, and C.~G. Lacey, {\it {The assembly bias of dark
  matter haloes to higher orders}},  {\em Mon. Not. Roy. Astron. Soc.} {\bf
  387} (2008) 921, [\href{http://arxiv.org/abs/0712.2280}{{\tt
  arXiv:0712.2280}}].

\bibitem{Dalal:2008}
N.~Dalal, M.~White, J.~R. Bond, and A.~Shirokov, {\it {Halo Assembly Bias in
  Hierarchical Structure Formation}},  {\em Astrophys. J.} {\bf 687} (2008)
  12--21, [\href{http://arxiv.org/abs/0803.3453}{{\tt arXiv:0803.3453}}].

\bibitem{Faltenbacher:2009}
A.~Faltenbacher and S.~D.~M. White, {\it {Assembly bias and the dynamical
  structure of dark matter halos}},  {\em Astrophys. J.} {\bf 708} (2010)
  469--473, [\href{http://arxiv.org/abs/0909.4302}{{\tt arXiv:0909.4302}}].

\bibitem{Sunayama:2015}
T.~Sunayama, A.~P. Hearin, N.~Padmanabhan, and A.~Leauthaud, {\it {The
  Scale-Dependence of Halo Assembly Bias}},  {\em Mon. Not. Roy. Astron. Soc.}
  {\bf 458} (2016), no.~2 1510--1516,
  [\href{http://arxiv.org/abs/1509.06417}{{\tt arXiv:1509.06417}}].

\bibitem{Miyatake:2015}
H.~Miyatake, S.~More, M.~Takada, D.~N. Spergel, R.~Mandelbaum, E.~S. Rykoff,
  and E.~Rozo, {\it {Evidence of Halo Assembly Bias in Massive Clusters}},
  {\em Phys. Rev. Lett.} {\bf 116} (2016), no.~4 041301,
  [\href{http://arxiv.org/abs/1506.06135}{{\tt arXiv:1506.06135}}].

\bibitem{More:2016}
S.~More et~al., {\it {Detection of the Splashback Radius and Halo Assembly bias
  of Massive Galaxy Clusters}},  {\em Astrophys. J.} {\bf 825} (2016), no.~1
  39, [\href{http://arxiv.org/abs/1601.06063}{{\tt arXiv:1601.06063}}].

\bibitem{Navarro:1996}
J.~F. Navarro, C.~S. Frenk, and S.~D.~M. White, {\it {A Universal density
  profile from hierarchical clustering}},  {\em Astrophys. J.} {\bf 490} (1997)
  493--508, [\href{http://arxiv.org/abs/astro-ph/9611107}{{\tt
  astro-ph/9611107}}].

\bibitem{Wechsler:2001}
R.~H. Wechsler, J.~S. Bullock, J.~R. Primack, A.~V. Kravtsov, and A.~Dekel,
  {\it {Concentrations of dark halos from their assembly histories}},  {\em
  Astrophys. J.} {\bf 568} (2002) 52--70,
  [\href{http://arxiv.org/abs/astro-ph/0108151}{{\tt astro-ph/0108151}}].

\bibitem{Neto:2007}
A.~F. Neto, L.~Gao, P.~Bett, S.~Cole, J.~F. Navarro, C.~S. Frenk, S.~D.~M.
  White, V.~Springel, and A.~Jenkins, {\it {The statistics of lambda CDM Halo
  Concentrations}},  {\em Mon. Not. Roy. Astron. Soc.} {\bf 381} (2007)
  1450--1462, [\href{http://arxiv.org/abs/0706.2919}{{\tt arXiv:0706.2919}}].

\bibitem{Borzyszkowski:2016}
M.~Borzyszkowski, C.~Porciani, E.~Romano-Diaz, and E.~Garaldi, {\it {ZOMG I:
  How the cosmic web inhibits halo growth and generates assembly bias}},
  \href{http://arxiv.org/abs/1610.04231}{{\tt arXiv:1610.04231}}.

\bibitem{Lee:2016}
C.~T. {Lee}, J.~R. {Primack}, P.~{Behroozi}, A.~{Rodriguez-Puebla},
  D.~{Hellinger}, and A.~{Dekel}, {\it {Properties of Dark Matter Halos as a
  Function of Local Environment Density}},  {\em ArXiv e-prints} (Oct., 2016)
  [\href{http://arxiv.org/abs/1610.02108}{{\tt arXiv:1610.02108}}].

\bibitem{Ludlow:2016}
A.~D. Ludlow, S.~Bose, R.~E. Angulo, L.~Wang, W.~A. Hellwing, J.~F. Navarro,
  S.~Cole, and C.~S. Frenk, {\it {The Mass-Concentration-Redshift Relation of
  Cold and Warm Dark Matter Halos}},  {\em Mon. Not. Roy. Astron. Soc.} {\bf
  460} (2016), no.~2 1214--1232, [\href{http://arxiv.org/abs/1601.02624}{{\tt
  arXiv:1601.02624}}].

\bibitem{Yang:2005}
X.~Yang, H.~J. Mo, and F.~C. van~den Bosch, {\it {Observational evidence for an
  age dependence of halo bias}},  {\em Astrophys. J.} {\bf 638} (2006)
  L55--L58, [\href{http://arxiv.org/abs/astro-ph/0509626}{{\tt
  astro-ph/0509626}}].

\bibitem{Tinker:2012}
J.~L. Tinker, M.~R. George, A.~Leauthaud, K.~Bundy, A.~Finoguenov, R.~Massey,
  J.~Rhodes, and R.~H. Wechsler, {\it {The Correlated Formation Histories of
  Massive Galaxies and Their Dark Matter Halos}},  {\em Astrophys. J.} {\bf
  755} (2012) L5, [\href{http://arxiv.org/abs/1205.4245}{{\tt
  arXiv:1205.4245}}].

\bibitem{Wang:2013}
L.~Wang, S.~M. Weinmann, G.~De~Lucia, and X.~Yang, {\it {Detection of galaxy
  assembly bias}},  {\em Mon. Not. Roy. Astron. Soc.} {\bf 433} (2013) 515,
  [\href{http://arxiv.org/abs/1305.0350}{{\tt arXiv:1305.0350}}].

\bibitem{Lin:2015}
Y.-T. Lin, R.~Mandelbaum, Y.-H. Huang, H.-J. Huang, N.~Dalal, B.~Diemer, H.-Y.
  Jian, and A.~Kravtsov, {\it {On Detecting Halo Assembly Bias with Galaxy
  Populations}},  {\em Astrophys. J.} {\bf 819} (2016) 119,
  [\href{http://arxiv.org/abs/1504.07632}{{\tt arXiv:1504.07632}}].

\bibitem{Zu:2016}
Y.~Zu, R.~Mandelbaum, M.~Simet, E.~Rozo, and E.~S. Rykoff, {\it {On the Level
  of Cluster Assembly Bias in SDSS}},
  \href{http://arxiv.org/abs/1611.00366}{{\tt arXiv:1611.00366}}.

\bibitem{Zentner:2016}
A.~R. {Zentner}, A.~{Hearin}, F.~C. {van den Bosch}, J.~U. {Lange}, and
  A.~{Villarreal}, {\it {Constraints on Assembly Bias from Galaxy Clustering}},
   {\em ArXiv e-prints} (June, 2016)
  [\href{http://arxiv.org/abs/1606.07817}{{\tt arXiv:1606.07817}}].

\bibitem{Zentner:2013}
A.~R. Zentner, A.~P. Hearin, and F.~C. van~den Bosch, {\it {Galaxy assembly
  bias: a significant source of systematic error in the galaxy–halo
  relationship}},  {\em Mon. Not. Roy. Astron. Soc.} {\bf 443} (2014), no.~4
  3044--3067, [\href{http://arxiv.org/abs/1311.1818}{{\tt arXiv:1311.1818}}].

\bibitem{Hearin:2015}
A.~P. Hearin, A.~R. Zentner, F.~C. van~den Bosch, D.~Campbell, and E.~Tollerud,
  {\it {Introducing decorated HODs: modelling assembly bias in the
  galaxy–halo connection}},  {\em Mon. Not. Roy. Astron. Soc.} {\bf 460}
  (2016), no.~3 2552--2570, [\href{http://arxiv.org/abs/1512.03050}{{\tt
  arXiv:1512.03050}}].

\bibitem{Chaves-Montero:2015}
J.~Chaves-Montero, R.~E. Angulo, J.~Schaye, M.~Schaller, R.~A. Crain,
  M.~Furlong, and T.~Theuns, {\it {Subhalo abundance matching and assembly bias
  in the EAGLE simulation}},  {\em Mon. Not. Roy. Astron. Soc.} {\bf 460}
  (2016), no.~3 3100--3118, [\href{http://arxiv.org/abs/1507.01948}{{\tt
  arXiv:1507.01948}}].

\bibitem{Lehmann:2015}
B.~V. Lehmann, Y.-Y. Mao, M.~R. Becker, S.~W. Skillman, and R.~H. Wechsler,
  {\it {The Concentration Dependence of the Galaxy-Halo Connection: Modeling
  Assembly Bias with Abundance Matching}},
  \href{http://arxiv.org/abs/1510.05651}{{\tt arXiv:1510.05651}}.

\bibitem{lemaitre:1933}
G.~{Lema{\^i}tre}, {\it {L'Univers en expansion}},  {\em Annales de la Societe
  Scietifique de Bruxelles} {\bf 53} (1933) 51.

\bibitem{sirko:2005}
E.~Sirko, {\it {Initial conditions to cosmological N-body simulations, or how
  to run an ensemble of simulations}},  {\em \apj} {\bf 634} (2005) 728--743,
  [\href{http://arxiv.org/abs/astro-ph/0503106}{{\tt astro-ph/0503106}}].

\bibitem{baldauf/etal:2011}
T.~Baldauf, U.~Seljak, L.~Senatore, and M.~Zaldarriaga, {\it {Galaxy Bias and
  non-Linear Structure Formation in General Relativity}},  {\em JCAP} {\bf
  1110} (2011) 031, [\href{http://arxiv.org/abs/1106.5507}{{\tt
  arXiv:1106.5507}}].

\bibitem{li/hu/takada:2014}
Y.~{Li}, W.~{Hu}, and M.~{Takada}, {\it {Super-sample covariance in
  simulations}},  {\em \prd} {\bf 89} (Apr., 2014) 083519,
  [\href{http://arxiv.org/abs/1401.0385}{{\tt arXiv:1401.0385}}].

\bibitem{Wagner:2014}
C.~Wagner, F.~Schmidt, C.-T. Chiang, and E.~Komatsu, {\it {Separate Universe
  Simulations}},  {\em Mon.Not.Roy.Astron.Soc.} {\bf 448} (2015) 11,
  [\href{http://arxiv.org/abs/1409.6294}{{\tt arXiv:1409.6294}}].

\bibitem{baldauf/etal:11}
T.~Baldauf, U.~Seljak, L.~Senatore, and M.~Zaldarriaga, {\it {Galaxy Bias and
  non-Linear Structure Formation in General Relativity}},  {\em JCAP} {\bf
  1110} (2011) 031, [\href{http://arxiv.org/abs/1106.5507}{{\tt
  arXiv:1106.5507}}].

\bibitem{jeong/etal}
D.~{Jeong}, F.~{Schmidt}, and C.~M. {Hirata}, {\it {Large-scale clustering of
  galaxies in general relativity}},  {\em \prd} {\bf 85} (Jan., 2012) 023504,
  [\href{http://arxiv.org/abs/1107.5427}{{\tt arXiv:1107.5427}}].

\bibitem{kaiser:1984}
N.~{Kaiser}, {\it {On the spatial correlations of Abell clusters}},  {\em
  \apjl} {\bf 284} (Sept., 1984) L9--L12.

\bibitem{Paranjape:2016}
A.~Paranjape and N.~Padmanabhan, {\it {Halo assembly bias from Separate
  Universe simulations}},  \href{http://arxiv.org/abs/1612.02833}{{\tt
  arXiv:1612.02833}}.

\bibitem{Zentner:2006}
A.~R. Zentner, {\it {The Excursion Set Theory of Halo Mass Functions, Halo
  Clustering, and Halo Growth}},  {\em Int. J. Mod. Phys.} {\bf D16} (2007)
  763--816, [\href{http://arxiv.org/abs/astro-ph/0611454}{{\tt
  astro-ph/0611454}}].

\bibitem{Musso:2012}
M.~Musso and R.~K. Sheth, {\it {One step beyond: The excursion set approach
  with correlated steps}},  {\em Mon. Not. Roy. Astron. Soc.} {\bf 423} (2012)
  L102--L106, [\href{http://arxiv.org/abs/1201.3876}{{\tt arXiv:1201.3876}}].

\bibitem{Paranjape:2012}
A.~{Paranjape} and R.~K. {Sheth}, {\it {Peaks theory and the excursion set
  approach}},  {\em \mnras} {\bf 426} (Nov., 2012) 2789--2796,
  [\href{http://arxiv.org/abs/1206.3506}{{\tt arXiv:1206.3506}}].

\bibitem{Paranjape:2013}
A.~{Paranjape}, R.~K. {Sheth}, and V.~{Desjacques}, {\it {Excursion set peaks:
  a self-consistent model of dark halo abundances and clustering}},  {\em
  \mnras} {\bf 431} (May, 2013) 1503--1512,
  [\href{http://arxiv.org/abs/1210.1483}{{\tt arXiv:1210.1483}}].

\bibitem{Lazeyras:2015}
T.~Lazeyras, C.~Wagner, T.~Baldauf, and F.~Schmidt, {\it {Precision measurement
  of the local bias of dark matter halos}},  {\em JCAP} {\bf 1602} (2016),
  no.~02 018, [\href{http://arxiv.org/abs/1511.01096}{{\tt arXiv:1511.01096}}].

\bibitem{Musso:2014}
M.~Musso and R.~K. Sheth, {\it {Stochasticity in halo formation and the
  excursion set approach}},  {\em Mon. Not. Roy. Astron. Soc.} {\bf 442}
  (2014), no.~1 401--405, [\href{http://arxiv.org/abs/1401.3185}{{\tt
  arXiv:1401.3185}}].

\bibitem{Robertson:2008}
B.~E. Robertson, A.~V. Kravtsov, J.~Tinker, and A.~R. Zentner, {\it {Collapse
  Barriers and Halo Abundance: Testing the Excursion Set Ansatz}},  {\em
  Astrophys. J.} {\bf 696} (2009) 636--652,
  [\href{http://arxiv.org/abs/0812.3148}{{\tt arXiv:0812.3148}}].

\bibitem{Castorina:2016}
E.~Castorina, A.~Paranjape, O.~Hahn, and R.~K. Sheth, {\it {Excursion set
  peaks: the role of shear}},  \href{http://arxiv.org/abs/1611.03619}{{\tt
  arXiv:1611.03619}}.

\bibitem{Bond:1993}
J.~R. Bond and S.~T. Myers, {\it {The Hierarchical peak patch picture of cosmic
  catalogs. 1. Algorithms}},  {\em Astrophys. J. Suppl.} {\bf 103} (1996) 1.

\bibitem{Monaco:1997}
P.~Monaco, {\it {A Lagrangian dynamical theory for the mass function of cosmic
  structures: 2. Statistics}},  {\em Mon. Not. Roy. Astron. Soc.} {\bf 290}
  (1997), no.~3 439--455, [\href{http://arxiv.org/abs/astro-ph/9606029}{{\tt
  astro-ph/9606029}}].

\bibitem{Pace:2014}
F.~Pace, R.~C. Batista, and A.~Del~Popolo, {\it {Effects of shear and rotation
  on the spherical collapse model for clustering dark energy}},  {\em Mon. Not.
  Roy. Astron. Soc.} {\bf 445} (2014), no.~1 648--659,
  [\href{http://arxiv.org/abs/1406.1448}{{\tt arXiv:1406.1448}}].

\bibitem{Sheth:1999}
R.~K. Sheth, H.~J. Mo, and G.~Tormen, {\it {Ellipsoidal collapse and an
  improved model for the number and spatial distribution of dark matter
  haloes}},  {\em Mon. Not. Roy. Astron. Soc.} {\bf 323} (2001) 1,
  [\href{http://arxiv.org/abs/astro-ph/9907024}{{\tt astro-ph/9907024}}].

\bibitem{Bardeen:1985}
J.~M. Bardeen, J.~R. Bond, N.~Kaiser, and A.~S. Szalay, {\it {The Statistics of
  Peaks of Gaussian Random Fields}},  {\em Astrophys. J.} {\bf 304} (1986)
  15--61.

\bibitem{Musso:2014b}
M.~Musso and R.~K. Sheth, {\it {On the Markovian assumption in the excursion
  set approach: the approximation of Markovian Velocities}},  {\em Mon. Not.
  Roy. Astron. Soc.} {\bf 443} (2014), no.~2 1601--1613,
  [\href{http://arxiv.org/abs/1401.8177}{{\tt arXiv:1401.8177}}].

\bibitem{Wagner:2015}
C.~Wagner, F.~Schmidt, C.-T. Chiang, and E.~Komatsu, {\it {The angle-averaged
  squeezed limit of nonlinear matter N-point functions}},
  \href{http://arxiv.org/abs/1503.03487}{{\tt arXiv:1503.03487}}.

\bibitem{Springel:2005}
V.~Springel, {\it {The Cosmological simulation code GADGET-2}},  {\em
  Mon.Not.Roy.Astron.Soc.} {\bf 364} (2005) 1105--1134,
  [\href{http://arxiv.org/abs/astro-ph/0505010}{{\tt astro-ph/0505010}}].

\bibitem{Gill:2004}
S.~P. Gill, A.~Knebe, and B.~K. Gibson, {\it {The Evolution substructure 1: A
  New identification method}},  {\em Mon.Not.Roy.Astron.Soc.} {\bf 351} (2004)
  399, [\href{http://arxiv.org/abs/astro-ph/0404258}{{\tt astro-ph/0404258}}].

\bibitem{Knollmann:2009}
S.~R. Knollmann and A.~Knebe, {\it {Ahf: Amiga's Halo Finder}},  {\em
  Astrophys.J.Suppl.} {\bf 182} (2009) 608--624,
  [\href{http://arxiv.org/abs/0904.3662}{{\tt arXiv:0904.3662}}].

\bibitem{Prada:2011}
F.~Prada, A.~A. Klypin, A.~J. Cuesta, J.~E. Betancort-Rijo, and J.~Primack,
  {\it {Halo concentrations in the standard LCDM cosmology}},  {\em Mon. Not.
  Roy. Astron. Soc.} {\bf 423} (2012) 3018--3030,
  [\href{http://arxiv.org/abs/1104.5130}{{\tt arXiv:1104.5130}}].

\bibitem{Bullock:2000}
J.~S. Bullock, A.~Dekel, T.~S. Kolatt, A.~V. Kravtsov, A.~A. Klypin,
  C.~Porciani, and J.~R. Primack, {\it {A Universal angular momentum profile
  for galactic halos}},  {\em Astrophys. J.} {\bf 555} (2001) 240--257,
  [\href{http://arxiv.org/abs/astro-ph/0011001}{{\tt astro-ph/0011001}}].

\bibitem{Davis:1985}
M.~Davis, G.~Efstathiou, C.~S. Frenk, and S.~D.~M. White, {\it {The Evolution
  of Large Scale Structure in a Universe Dominated by Cold Dark Matter}},  {\em
  Astrophys. J.} {\bf 292} (1985) 371--394.

\bibitem{Klypin:1999}
A.~{Klypin}, S.~{Gottl{\"o}ber}, A.~V. {Kravtsov}, and A.~M. {Khokhlov}, {\it
  {Galaxies in N-Body Simulations: Overcoming the Overmerging Problem}},  {\em
  \apj} {\bf 516} (May, 1999) 530--551,
  [\href{http://arxiv.org/abs/astro-ph/9708191}{{\tt astro-ph/9708191}}].

\bibitem{Skibba:2011}
R.~A. {Skibba} and A.~V. {Macci{\`o}}, {\it {Properties of dark matter haloes
  and their correlations: the lesson from principal component analysis}},  {\em
  \mnras} {\bf 416} (Sept., 2011) 2388--2400,
  [\href{http://arxiv.org/abs/1103.1641}{{\tt arXiv:1103.1641}}].

\end{thebibliography}\endgroup
\end{document}